\documentclass[trackchanges]{aastex701}

\def\Rs{R_{\odot}}

\def\thprime{\Theta^{\prime}}

\newcommand{\mean}[1]{\overline{#1}}
\newcommand{\brac}[1]{\langle #1 \rangle}
\usepackage{hyperref}
\usepackage[utf8]{inputenc}
\usepackage{graphicx}
\usepackage{amsmath}	

\usepackage{graphics}

\begin{document}
\title{Simulations of the onset and dynamical evolution of inertial waves in solar/stellar interior}

\author[orcid=0009-0000-3220-5535]{Souza-Gomes, M. D.,}
\affiliation{Physics Department, Universidade Federal de Minas Gerais, Av. Antonio Carlos, 6627, Belo Horizonte, MG 31270-901, Brazil}
\affiliation{Observatório Nacional, Rua General José Cristino 77, CEP 20921-400, São Cristóvão, Rio de Janeiro, RJ, Brazil}
\email[show]{marianegomes@on.br}  

\author[orcid=0009-0004-8465-3881]{Finotti, C.S.} 
\affiliation{Physics Department, Universidade Federal de Minas Gerais, Av. Antonio Carlos, 6627, Belo Horizonte, MG 31270-901, Brazil}
\email{conradofinotti@ufmg.br}

\author[orcid=0000-0002-2671-8796]{Guerrero, G.} 
\affiliation{Physics Department, Universidade Federal de Minas Gerais, Av. Antonio Carlos, 6627, Belo Horizonte, MG 31270-901, Brazil}
\email{guerrero@fisica.ufmg.br}

\author[orcid=0000-0002-7679-3962]{Triana, S. A.} 
\affiliation{Royal Observatory of Belgium, Ringlaan 3, B-1180 Brussels, Belgium}
\email{santiago.triana@observatory.be }

\author[orcid=0000-0002-2227-0488]{Dikpati, M.} 
\affiliation{High Altitude Observatory, NSF  National Center for Atmospheric Research, Boulder, Colorado, USA}
\email{dikpati@ucar.edu}

\author[orcid=0000-0001-7077-3285]{Smolarkiewicz, P.K.} 
\affiliation{ National Center for Atmospheric Research, Boulder, Colorado, USA}
\email{smolar@ucar.edu}

\author[orcid=0009-0005-3657-8183]{Botelho, E.S.} 
\affiliation{Physics Department, Universidade Federal de Minas Gerais, Av. Antonio Carlos, 6627, Belo Horizonte, MG 31270-901, Brazil}
\email{ericsobo@ufmg.br}

\begin{abstract}
Inertial modes have been recently detected in the Sun via helioseismology, yet their origin, evolution, and role in the dynamics of the solar plasma and magnetic field remain poorly understood. In this study, we employ global numerical simulations to investigate the excitation mechanisms and dynamical consequences of inertial modes in the Sun and stellar interiors. We validate first our numerical setup by analyzing the evolution of sectoral and tesseral perturbations imposed on a rigidly rotating sphere. The results confirm that a perturbation of a given mode can excite neighboring modes with both smaller and larger wavenumbers along the dispersion relation of Rossby waves.
Subsequently, we use a physically motivated forcing to impose differential rotation with varying shear amplitudes, and examine the spontaneous onset and nonlinear evolution of inertial modes. The simulations reveal that the growth of velocity perturbations is primarily driven by baroclinic instability. It gives rise to high-latitude inertial modes in the form of retrograde polar vortices whose properties depend on the imposed shear. Equatorial Rossby modes are also excited, albeit with lower intensity than their high-latitude counterpart. Perturbations with arbitrary azimuthal wavenumbers lead to the excitation of Rossby modes for all available wave numbers, sustained by both direct and inverse energy cascades. In simulations with stronger shear, the high latitude modes produce Reynolds stresses able to modify the imposed differential rotation and accelerate the rotation of the poles.
\end{abstract}


\section{Introduction}

Perturbations in a stably or unstably stratified differentially rotating gas subject to gravity and rotation restoring forces give rise to a plethora of waves and instabilities. Examples of the latter are the convective,  shear, baroclinic and barotropic instabilities, examples of waves are the internal gravity waves and various inertial modes including the famous Rossby waves. 
In the Earth context, both, instabilities and waves \citep[e.g.,][]{rossby1939planetary,1940TrAGU..21..262H} have been vastly documented and studied.  For the case of astrophysical bodies we have mainly theoretical and numerical predictions, yet not sufficient observational evidence \citep[see][for a review]{2021SSRv..217...15Z}.  Starting from \cite{loptien2018global}, solar observations identified various inertial patterns in the Sun's surface through helioseismic \citep{hanasoge2019detection, liang2019time,proxauf2020exploring,hanson2020solar,gizon2021solar} and motion tracking techniques \citep{2021ApJ...908..160H}. Nowadays,
the existence of the equatorial Rossby waves, the high latitude modes, the critical layer inertial modes, and the high frequency modes reported by \cite{Hanson2022NatAs} is widely accepted.

These distinct inertial modes exhibit characteristics that make them key probes of the solar interior’s dynamics. Equatorial Rossby modes are large-scale, predominantly horizontal oscillations that propagate retrograde in the rotating frame and follow the dispersion relation $\omega = -2m\Omega / l(l+1)$, where $l$ and $m$ are the angular degree and azimuthal wavenumber, respectively. In contrast, the vorticity of the high-latitude modes manifest as spiral patterns near the poles and exhibit a strong sensitivity to latitudinal entropy gradients and differential rotation \citep{Fournier2022A6,bekki2022theory}. Recent work indicates that, at low azimuthal order, these modes may become baroclinically unstable, potentially contributing to the transport of angular momentum and, therefore, regulating the solar differential rotation \citep{bekki2024sun}.
The Sun’s latitudinal differential rotation (DR) introduces critical layers at latitudes where the phase velocity of a mode matches the local rotation rate, yielding singularities in the inviscid eigenvalue problem (e.g. \cite{1981GApFD..16..285W, 1999ApJ...526..523C}). In a viscous fluid, these singularities are regularized, giving rise to quasi-toroidal inertial oscillations with energy concentrated near these critical layers. \cite{gizon2021solar} refers to these modes as \emph{critical-latitude} modes.  While this denomination is motivated by the presence of critical layers, as explained above, it may be potentially misleading  since it is not directly related  to the \emph{critical latitudes} of inertial oscillations in uniformly rotating fluids which are well established in the literature 
\citep{Greenspan1968,2001JFM...435..103R}. Finally, the high-frequency modes exhibit antisymmetric vorticity about the equator and possess a frequency approximately three times higher than that of sectoral Rossby modes \citep{hanson2022discovery}.

In the stellar context, non-radial oscillations have long been the subject of theoretical investigation. The inclusion of rotation in the eigenvalue problem was examined by \cite{papaloizou1978non}, and subsequently extended by \cite{dziembowski1987low} and \cite{dziembowski1987lowII}, who incorporated differential rotation and applied the formalism to solar models.
These studies found  that the Rossby modes may develop in cavities either below the convection zone or near the photosphere. In these locations the ratio $N^2/\Omega_0^2$ is above unity; here $N$ marks the Brunt-Väisälä frequency and $\Omega_0$ is the angular velocity of the rotating frame. They also explored the development of oscillations driven by the Kelvin-Helmholtz instability. Depending on the shear profile, these oscillations may destabilize the DR. Further developments on these shear instabilities were performed by \cite{dikpati2001analysis}.

More recent theoretical studies have focused on reproducing the solar observations, most of them solving the linear eigenvalue problem. These include studies of the critical layer modes \citep{gizon2020effect} and the equatorial Rossby modes \citep{bekki2022theory, 2023ApJS..264...21B}. Furthermore, \cite{triana2022identification} demonstrated that the eigenvalue solution for a solid rotating cavity is able to capture both, the equatorial Rossby modes and the high frequency modes. These modes were also reproduced by \cite{2023ApJS..264...21B}, and by  \cite{2024A&A...682A..39B} who explored the dependence of the high frequency modes on the DR and the adiabaticity of the layer. 
Extending beyond the linear framework, in the nonlinear regime \citet{2022A&A...666A.135B} identified equatorial Rossby modes together with other inertial modes, including high-latitude modes, the former being stochastically excited by convection and reaching amplitudes comparable to those observed on the Sun.
In other words, these works have focused on the excitation mechanisms of inertial modes.

Other efforts have focused on identifying inertial modes in global convective dynamo simulations \citep{2024ApJ...966...29B}. That study demonstrated that inertial modes can coexist with convective and dynamo instabilities. However, because of the complexity of the simulation, the study did not clarify the origin of the modes nor their role in the overall dynamics of the simulations. More recently, \cite{bekki2024sun} addressed the origin of the high-latitude modes using a 3D mean-field model of differential rotation based on the theory of the $\Lambda$-effect \citep[see][and references therein]{2005ApJ...622.1320R} that permits non-axisymmetric perturbations and employs a modified continuity equation derived from the reduced speed of sound approximation. Their results suggest that the high-latitude modes are driven by baroclinic instability, and that the properties of these modes are sensitive to the thermal stratification at the base of the convection zone.  

Despite the remarkable progress made in recent years, theoretical investigations still lack an integrated framework that simultaneously addresses the conditions in the solar interior conducive to the excitation of inertial modes and their subsequent nonlinear evolution. Furthermore, the contribution of these modes to the overall dynamics of solar and stellar interiors still requires a better understanding.

The objective of the present work is to fill this gap by developing a numerical model that reproduces the conditions of solar and stellar interiors and enables the investigation of the onset and nonlinear evolution of inertial modes under these conditions. The model is implemented using the hydrodynamic branch of the EULAG-MHD code \citep{smolarkiewicz2013eulag}. We first validate our model by studying the evolution of Rossby waves in rigidly rotating layers. We then introduce DR by adding a body-force term to the momentum equation in such a way that the longitudinal velocity relaxes to the target mean flow over an arbitrary time scale. Yet, we allow for the development of non-axysymmetric perturbations along the mean flow, evolving in fluid layers with thermodynamic stratification representative of the solar interior while exploring DR profiles resembling the solar case yet with different amount of shear. We examine the growth of these perturbations as well as their spectral structure and their spatial and temporal propagation. 

This paper is organized as follows. In Section~\ref{s:model}, we describe the numerical model. In Section~\ref{s:results}, we first present the results for solid-body rotation, which serve to validate the model. We then examine the case with differential rotation, interpret the results in the framework of the Taylor–Proudman theorem, propose the instability most relevant for the development of inertial modes, and analyze the evolution of these modes and their contribution to angular momentum transport. Finally, in Section~\ref{s:conclusions}, we summarize our conclusions.

\section{Numerical model}
\label{s:model}

To study the non-linear dynamics of inertial modes we use the hydrodynamic modules of the EULAG-MHD code. It solves the Lipps-Hemler anelastic system of equations \citep{LippsHemler1982,Lipps1990},
\begin{eqnarray}
\boldsymbol{\nabla} \cdot \left(\rho_{\rm ad} \boldsymbol{u}\right) &=& 0 \;, \label{eq:mass} \\
\frac{D\boldsymbol{u}}{Dt}+2\boldsymbol{\Omega}\times\boldsymbol{u} &=& -\boldsymbol{\nabla}\left(\frac{p'}{\rho_{\rm ad}}\right)+\boldsymbol{g}\frac{\Theta^\prime}{\Theta_{\rm ad}} - \boldsymbol{F} \; , \label{eq:mom} \\
\frac{D\Theta^\prime}{Dt}&=&-\boldsymbol{u}\cdot\boldsymbol{\nabla}\Theta_{\rm amb}-\frac{\Theta^\prime}{\tau}, \label{eq:pott}
\end{eqnarray}

where $D/Dt=\partial/\partial t + \boldsymbol{u}\cdot\boldsymbol{\nabla}$ is the material derivative,

$\boldsymbol{u} = (u,v,w)$ is the velocity field in a rotating frame with $\boldsymbol{\Omega} =\Omega_0(cos\theta,-sin\theta, 0)$, where $\Omega_0$
is the angular velocity. 
The pressure-perturbation variable, $p'$, accounts for the gas pressure while subsuming the centrifugal force. The potential temperature perturbation, $\Theta^\prime$, enters both the buoyancy term in the momentum Eq.~\ref{eq:mom} and the potential temperature, Eq.~\ref{eq:pott}; $\Theta$ is related to the specific entropy via $s = c_p \ln\Theta + {\rm const}$, where $c_p$ is the specific heat. The $\Theta^\prime$ perturbations are defined with respect to a presumed ambient state, $\Theta_{\rm amb}$, considered a known truth of stellar interiors based on the stellar structure models. The last term on the rhs of Eq.~\ref{eq:pott} relaxes $\Theta^\prime$ to zero in a  timescale $\tau = 1.3 \times  10^7\mathrm{s}$. The existence of these perturbations is fundamental for the model to achieve a relaxed equilibrium state. The value of $\tau$ corresponds to 5 solar rotations, which is a numerically convenient compromise between capturing essential physics and ensuring a rapid thermal relaxation. The variables $\rho_{\rm ad}$ and $\Theta_{\rm ad}$ are, respectively, the density and potential
temperature of the hydrostatic adiabatic reference state, whereas $\boldsymbol{g} = g \hat{\bf e}_r$ is the gravity acceleration of the solar interior adjusted from the solar model of \citet{christensen1996current}. The last term in Eq.~\ref{eq:mom}, $\boldsymbol{F}$, is a physically motivated body-force acting on the longitudinal component of the velocity field as will be discussed below.

The adiabatic and ambient states required for the thermodynamic variables, hereafter, $\rho_{\rm ad}$, $\Theta_{\rm ad}$, and $\Theta_{\rm amb}$, are built considering polytropic equations of state, $p \propto \rho^{1+1/{m_p}_i}$, for the hydrostatic problem. This results in the following equations,
\begin{equation}\label{eq:temp_amb}
    \centering
    \frac{\partial T_i}{\partial r} = - \frac{g}{R_g\left({m_p}_i+1\right)};
\end{equation}
\begin{equation}\label{eq:density_amb}
\centering
\frac{\partial\rho_i}{\partial r} = -\frac{\rho_i}{T_i}\left(\frac{g}{R_g}+\frac{\partial T_i}{\partial r}\right).
\end{equation}
where the index $i$ stands for ``ad" (adiabatic) or ``amb" (ambient),  $R_g = 13732$ is the gas constant of a monatomic hydrogen gas, and ${m_p}_i$ is the polytropic index.  

The equations (\ref{eq:mass}-\ref{eq:pott}) are solved in a spherical domain with $0\le \phi \le 2\pi$, $0\le \theta \le \pi$. For the simulations with solid body rotation we consider a radial domain spanning from $r_{\rm bot} = 0.60 \Rs$ to $r_{\rm top} = 0.75 \Rs$.  For the adiabatic state ${m_p}_{\rm ad} = 1.5$, whereas for the ambient state we consider two layers. From $r_{\rm bot}$ to $r_{\rm rc} = 0.7 \Rs$, the shell is stable to convection. Above $ 0.7 \Rs$ the layer is neutral, or adiabatic. The polytropic index is defined using an ${\rm erf}$ function,
\begin{equation}
\label{eq:politropic}    
    {m_p}_{\rm am} (r) = m_0 - \frac{1}{2}(m_0 - m_1)\left [ 1 + \mathrm{erf}\left ( \frac{r - r_{\rm rc}}{w} \right ) \right ] \; ,
\end{equation}
where $m_0=2.5$ and $m_1=1.5$, and $w=0.01 \Rs$. For this model, the last term on the rhs of Eq.~\ref{eq:mom} is set to zero, $F=0$. 
The results of the simulations performed with this model will be discussed in \S\ref{s:solid}.

\begin{figure}                 
\begin{center}
    \includegraphics[width=0.7\columnwidth]{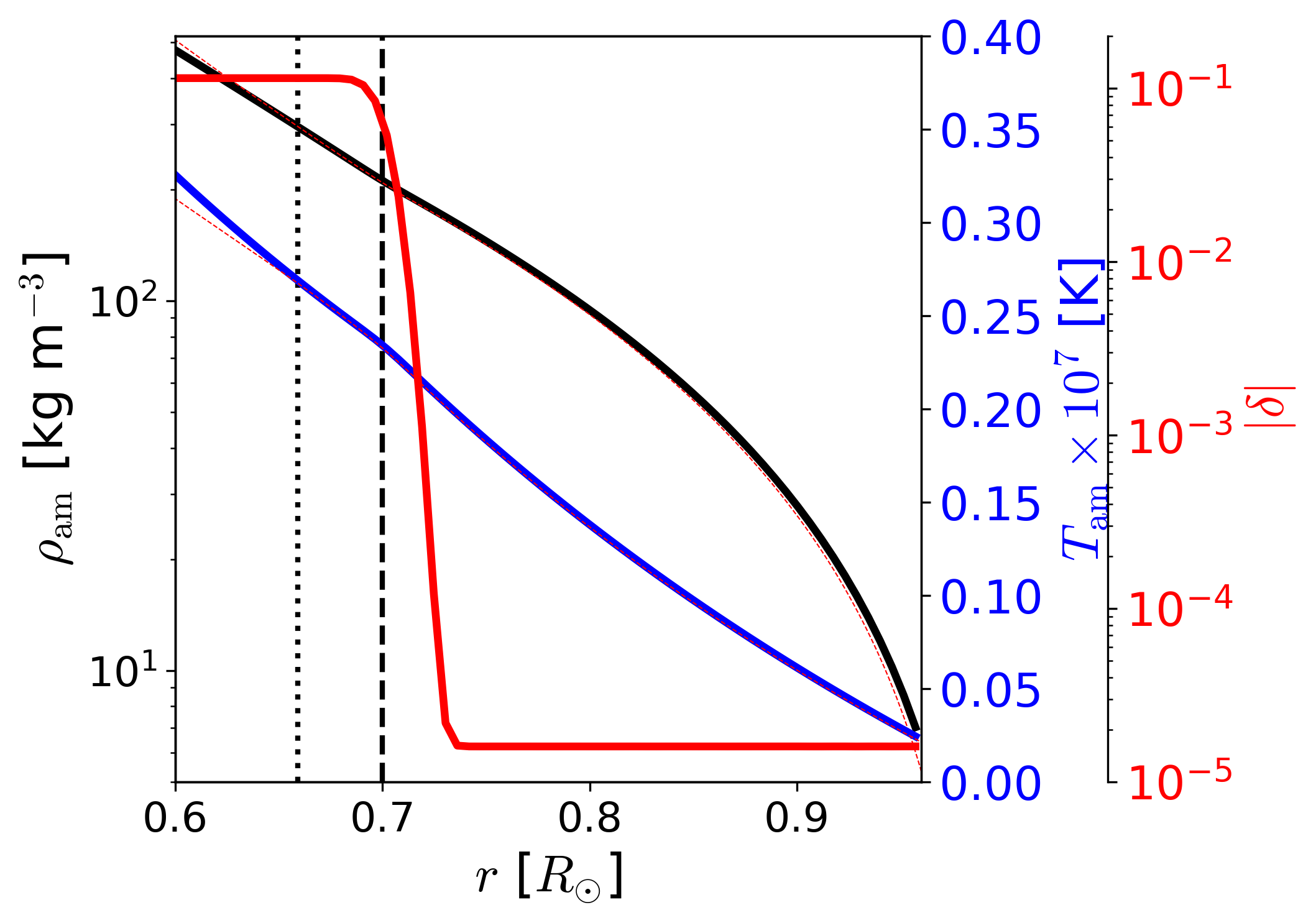}
    \caption{Thermodynamic structure of the simulations including shear. The black and blue lines depict the density and temperature of the ambient state as a function of radius. The thin red dashed line shows the same quantities for the solar model of \citep{christensen1996current}. The red line correspond to the adiabaticity at different layers(see the text). The vertical dashed and dotted lines correspond to the transition between a highly sub-adiabatic and a slightly sub-adiabatic layers, and the location of the tachocline, respectively.
} 
\label{fig:strat}
\end{center}
\end{figure}

For the simulations with differential rotation we consider a thicker domain, with $r_{\rm top}=0.95 \Rs$. To explore the behavior of inertial waves in cavities with different properties, we aim having a differential rotation profile,  including the strong radial shear of the tachocline, divided in three layers. The bottom layer has a stiff stratification, and rotates as a solid body.  The second layer has a stiff stratification and rotates differentially. The upper layer is adiabatic (or slightly subadiabatic, see \S\ref{s:TP}) and rotates differentially. 
To construct this model, we adopt an ambient state described by Eq.~(\ref{eq:politropic}) with $m_0=2.5$ and $m_1=1.5$ (or $m_1=1.5001$, see \S\ref{s:TP}). The resulting density and temperature profiles are shown in Fig.~\ref{fig:strat}. The thin red lines correspond to the solar model of \cite{christensen1996current}, illustrating that our stratification is well constrained by a widely accepted solar reference model. The thick red line shows the absolute value of the adiabaticity of the domain, $\delta = \nabla_{\rm amb} - \nabla_{\rm ad}$, with $\nabla = (p/T)\,(dT/dp)$. When $\delta > 0$ the system is Schwarzschild unstable, while $\delta < 0$ indicates stability. The case $\delta = 0$ corresponds to adiabatic stratification.  The vertical dashed and dotted lines correspond to the transition between a highly sub-adiabatic and a slightly sub-adiabatic layers, and the location of the tachocline, respectively.

Additionally, we consider the last term on the rhs of Eq.~\ref{eq:mom} as a body force acting on the longitudinal velocity, $u$, 
\begin{equation}
\label{eq:body_force}
        {\bf F} = F \hat{{\bf e}}_{\phi} = \frac{u - u_{F}}{\tau_s} \hat{{\bf e}}_{\phi} \; . 
\end{equation}
It relaxes $u$ to the target profile, $u_F$, on a time scale, $\tau_s$ = $1.29 \times 10^7$ s (unless stated otherwise), corresponding to roughly $5$ Carrington rotations.  

The target profile is a simple fit to the solar differential rotation, $u_{\rm F} = r \sin (\theta) \Omega_{\rm F}$, with
\begin{equation}
\label{eq:DR}
    \Omega_F = \left[\Omega_R - g(r)(\Omega_R - f(\theta)) \right] - \Omega_0 \; ,
\end{equation}
where $\Omega_R/2\pi =  432.8$ nHz is the rotation of the stable interior,
\begin{equation}
f(\theta)= \Omega_0 + \Omega_2 \cos^2\theta \;, 
\end{equation}
is the function for latitudinal differential rotation, and 
\begin{equation}
\label{eq:drerror}
        g(r) = \frac{1}{2} \left [ 1 + \mathrm{erf}\left ( \frac{r - r_{\rm tac}}{w_1} \right ) \right ] \; ,
\end{equation}
is the radial shear profile, where $r_{\rm tac} = 0.66 R_\odot$ is the radius of transition between solid body and latitudinal differential rotation, and  $w_1=0.01 \Rs$ is the thickness of such transition.  These parameters guarantee that the tachocline, and the strong radial shear, is located in a stiff radiative zone, as it has been revealed by 30 years of helioseismology inversions \citep{Basu_2025}.  With this profile of DR we force the equator to rotate at the same speed that the rotating frame, whereas the radiative interior rotates at the same speed that intermediate latitudes. This forcing term is used in the simulations presented in \S\ref{s:dr} where we explore the development and dynamics of inertial modes in the presence of shear. 

The velocity boundary conditions are impermeable and stress-free at both radial boundaries of the simulations. For the thermal boundary conditions null enthalpy radial ﬂux is implemented at the bottom and top. 

For the simulations with solid body rotation, the initial conditions for the latitudinal velocity correspond to spherical harmonics of different order. The initial condition for the other two components of the velocity, $u$ and $w$, as well as for the potential temperature perturbations, $\thprime$, is set to zero.  For the simulations with differential rotation,  the forcing, $F$, is turned on, and all other variables are set to zero.

\section{Results}
\label{s:results}
In the following sections we present the results of simulations performed with (i) solid body rotation and (ii) radial and latitudinal differential rotation imposed through the body force term $F$ (Eq.~\ref{eq:body_force}). 

\subsection{Inertial modes in simulations with solid body rotation}
\label{s:solid}

For the cases presented in this section we consider $(128,64,47)$ grid points resolution for the longitudinal, latitudinal and radial direction, respectively, the angular velocity of the rotating frame, $\Omega_0 = 2.8 \times 10^{-6}$ rad/s, equivalent to a rotation period,  $P_{rot} = 26$ earth days. The simulations were performed with an initial condition for the latitudinal component of the velocity vector, $v$, given by  normalized spherical harmonics,
\begin{equation*}
         v(\theta,\phi) = v_0\sqrt{\frac{(2l+1)(l-m)!}{4\pi(l+m)!}}P_{l}^{m}(\cos\theta)e^{im\phi} \; ,
\end{equation*}
where $v_0= 1$ m/s is the amplitude of the initial perturbation, $l$ and $m$ are, respectively, the latitudinal angular degree and the longitudinal wave number, and $P_{l}^{m}$ are the associated Legendre polynomials.

\begin{figure}                 
\begin{center}
    \includegraphics[width=0.7\columnwidth]{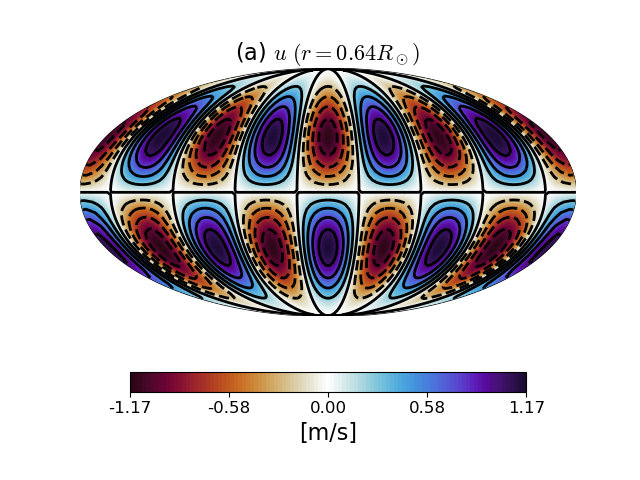}
    \includegraphics[width=0.7\columnwidth]{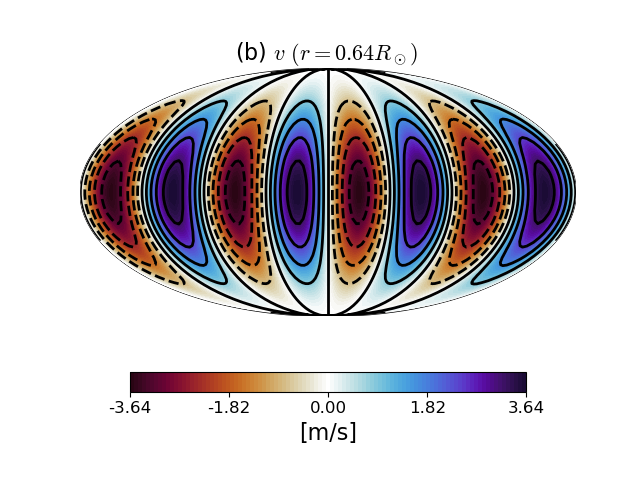}
    \caption{Mollweide projection of the velocity components $u$ (a) and $v$ (b) corresponding to the initial condition of the simulation with the sectoral mode $(m,l)=(4,4)$. Note that $u$ is anti-symmetric whereas $v$ is symmetric across the equator.}
    \label{fig:IC44}
\end{center}
\end{figure}

We performed simulations with both, sectoral and tesseral spherical harmonics as initial conditions. The symmetry of the modes depends on the character of the harmonic used in the perturbation. For sectoral modes, the longitudinal velocity $u$ is anti-symmetric across the equator and the latitudinal velocity $v$ is symmetric across the equator. For tesseral modes, this depends also on the difference $l-m$. If this number is even, we have the same symmetry of sectoral modes. If this number is odd, the longitudinal velocity $u$ is symmetric across the equator and the latitudinal velocity $v$ is anti-symmetric.  To assure solenoidality of the initial flow on
the model grid, the analytic initial condition for $v$ is supplemented with a gradient of potential $\phi$, to form the elliptic Poisson equation for $\phi$. The initial conditions are therefore not strictly pure spherical harmonics, although this has no practical impact on the results. The numerical solution of the resulting discrete problem (commonly referred to as the exact projection) satisfies the flow solenoidality on the grid to a specified precision. Technical details of the approach are widely documented in the EULAG literature; cf. \citep[][and references therein]{2016JCoPh.314..287S}. Figure~\ref{fig:IC44} shows a Mollweide projection of the initial condition for the velocity components $u$ and $v$ when the perturbation is a sectoral mode with $(m,l)=(4,4)$.

\begin{figure*}
\begin{center}
\includegraphics[width=\columnwidth]{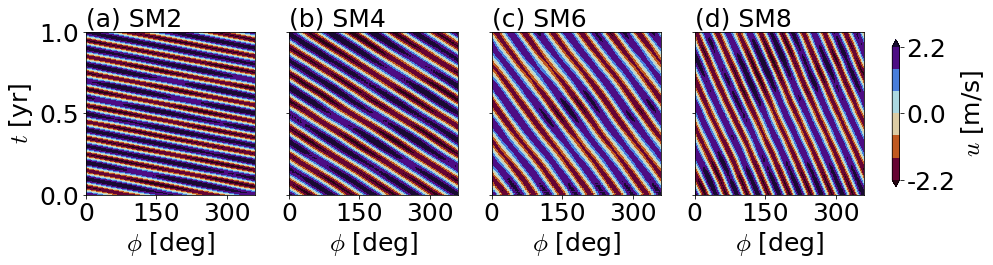}
\caption{Longitude-time, Hovmöller, diagrams showing the propagation of the longitudinal velocity, $u$ for simulation with sectoral initial condition $m=l=2,4,6,8$, panels (a) to (d), respectively, at a radius $r=0.64\Rs$ and $45^{\circ}$ latitude. The tilt of the propagating pattern depicts the phase speed of the wave. The thickness of the bands with alternating colors depicts the frequency of the mode.}
\label{fig:hovmoller_sectoral}
\end{center}
\end{figure*}

Once the simulation starts, the initial perturbation propagates in the retrograde direction. This is a consequence of the conservation of the total vorticity. The Coriolis force restores the parcels displaced in the latitudinal direction to the original position. When doing so, there is a difference in the planetary vorticity with latitude, which results in a longitudinal displacement, forming a wave pattern.  The propagation of these waves can be observed in Hovm\"oller diagrams \citep{hovmoller1949trough}, which have longitude in the abscissa and time in the ordinate axis and shows the value of $u$ at a particular radius and latitude.  Figure \ref{fig:hovmoller_sectoral} shows the Hovm\"oller diagrams for simulations with initial perturbation given by the sectoral modes $m=l=2,4,6,8$ (simulations SM2, SM4, SM6 and SM8, respectively), for $r=0.64 \Rs$ and $45^o$ latitude. The negative inclination indicates that the wave propagates in the opposite direction of the sphere's rotation, i.e. in the east-west direction. The tilt represents the phase speed of the wave and the thickness of the tilted bands (with red and blue colors) corresponds to the frequency of the wave. These propagating modes are the sectoral Rossby waves. 

Since the initial conditions do not have any radial variation and the modes evolve similarly in both, the bottom (sub-adiabatic) and top (adiabatic) layers, there is no significant change in the Hovm\"oller diagrams taken at different radii (not presented here for shortness). This can be observed in the animation corresponding to simulation SM4, available in \url{https://zenodo.org/records/18388135}.  From the Hovm\"oller diagrams it can be observed that there is a pulsation of the modes, i.e. there is a changing amplitude of the waves as they evolve in time. 
These pulses seem to arise from the interactions of waves evolving in two layers with different thermodynamical properties. 

\begin{figure*}[h!]
\begin{center}
\includegraphics[width=\columnwidth]{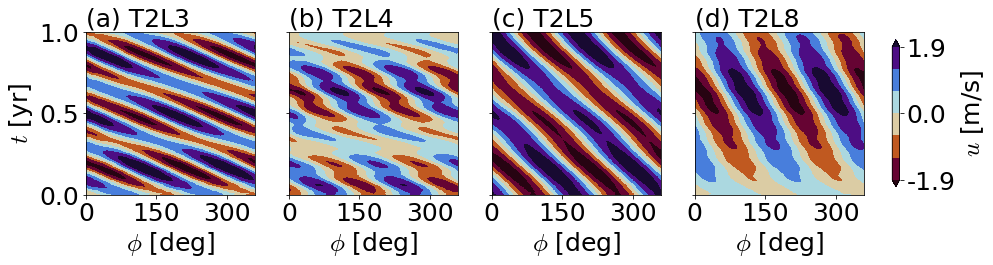}
\caption{Same as Fig.~\ref{fig:hovmoller_sectoral} but for tesseral initial perturbations with (a) $m=2$, $l=3$, (b) $m=2$, $l=4$, (c) $m=2$, $l=5$, and (d) $m=2$, $l=8$.}
\label{fig:hovmoller_tesseral}
\end{center}
\end{figure*}

The simulations with tesseral mode perturbations $m=2$, $l=3$ (simulation T2L3),  $m=2$, $l=4$ (T2L4),  $m=2$, $l=5$ (T2L5), and  $m=2$, $l=8$ (T2L8) are presented in Fig.~\ref{fig:hovmoller_tesseral} panels (a)-(d), respectively. Besides similar changes in phase speed and frequency, the non-linear evolution of these modes is in general more complex than for the sectoral modes.  The mode pulsation are more significant in this case. Also, in the Fig.~\ref{fig:hovmoller_tesseral}(b) it can be observed that as the waves evolve in the retrograde direction, they seem to stop and shortly become prograde, before continuing in their original direction. This can be related to the exchange of energy or/and cascade of modes to different wave numbers $m$. 

The modes in the lower (subadiabatic) layer seem preserve the radius independent structure of the initial perturbation. Conversely, the modes in the upper (adiabatic) layer acquire a radial profile, becoming aligned with the rotation axis and leading to the interaction of modes from different radius and latitudes \href{https://zenodo.org/records/18388135}{(click here to see the animation corresponding to simulations T2L5 and T2L8)}\footnote{All the animations are available in https://zenodo.org/records/18388135.}. This interaction of modes creates the complex patterns of evolution observed in Fig.~\ref{fig:hovmoller_tesseral}.

\begin{figure}
\begin{center}
    \includegraphics[width=0.45\columnwidth]{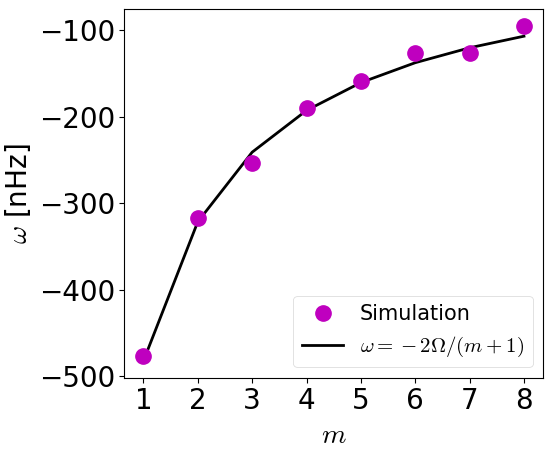}
    \includegraphics[width=0.45\columnwidth]{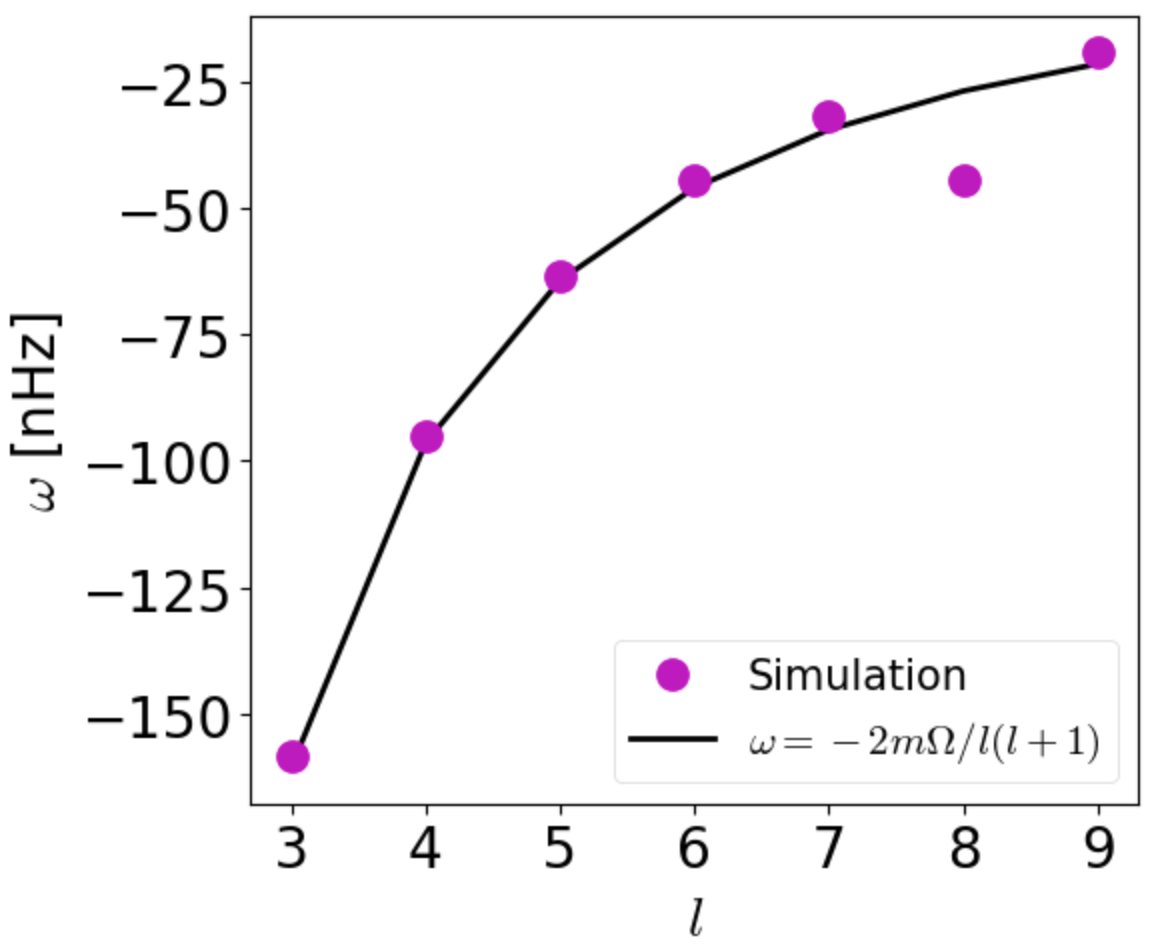}
    \caption{Wave frequency with highest power as a function of the longitudinal wave number, $m$ imposed as IC. The left panel corresponds to sectoral modes. The right panel corresponds to  tesseral modes with $m=2$ and different $l > m$.} The black lines show the theoretical dispersion relation, Eq.~(\ref{eq:tdr}).
    \label{fig:disp_relation}
\end{center}
\end{figure}

To confirm that these are inertial waves, we analyzed the power spectra of the longitudinal velocity of each model. For each simulation, the most energetic mode corresponds to that imposed by the initial perturbation. The frequency of this mode for simulations with different initial conditions can be seen in Fig.~\ref{fig:disp_relation}.  In general, the obtained frequencies match the theoretical dispersion relation obtained for perturbations evolving in time as $\cos(\omega t)$ \citep{1940TrAGU..21..262H},
\begin{equation}
    \omega = -2 m \frac{\Omega_0}{l(l+1)},
    \label{eq:tdr}
\end{equation}
for Rossby waves with sectoral (left panel) and tesseral modes (right panel). Note that tesseral Rossby modes fulfill Eq.~(\ref{eq:tdr}) only if radial motions are suppressed, as it is the case thanks to buoyancy forces in the radiative zone, or thanks to a radially thin convective zone (0.05 $R_\odot$ in this case), see \citet{triana2022identification}.  

\begin{figure*}[h!]
\begin{center}
\includegraphics[width=\columnwidth]{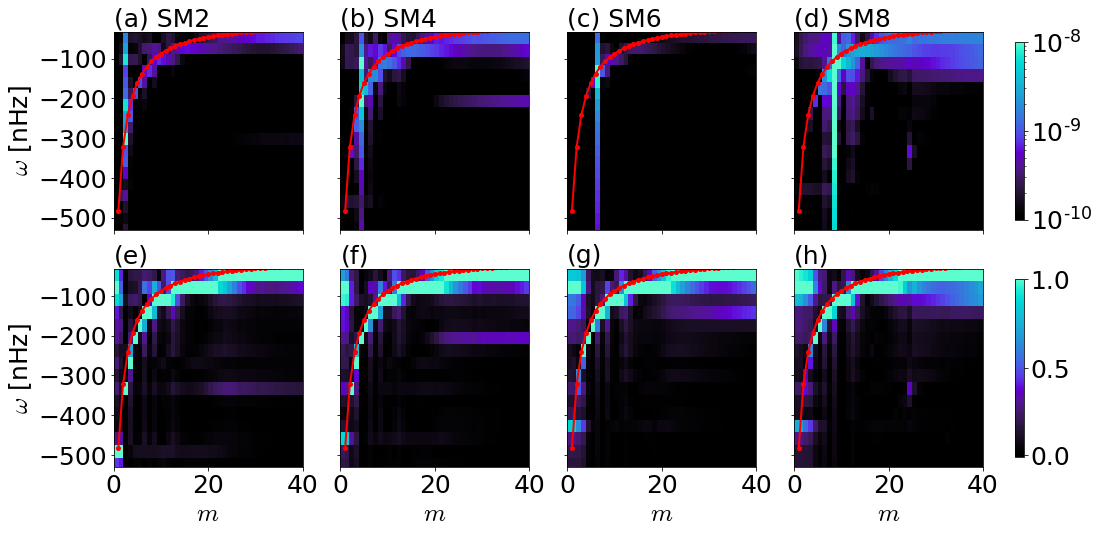} 
\caption{Power spectra of the radial vorticity, $\tilde{\omega_r}$, computed in a shell located at $r=0.64\Rs$, as a function of the longitudinal wave number, $m$, for the same simulations presented in Fig.~\ref{fig:hovmoller_sectoral}. The color intensity represent the strength of the mode. The red line depicts the theoretical dispersion relation, Eq.~\ref{eq:tdr}. The upper and bottom rows correspond to unnormalized and normalized power spectra. }
\label{fig:spectra_sec}
\end{center}
\end{figure*}

\begin{figure*}
\begin{center}
    \includegraphics[width=\columnwidth]{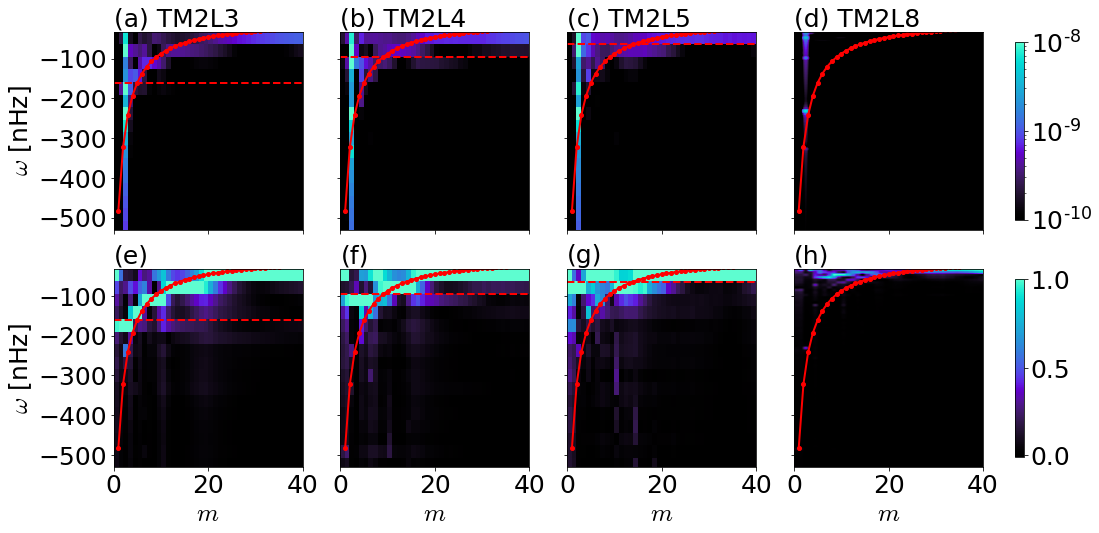} 
\caption{Same as Fig.~\ref{fig:spectra_sec} but for the simulations with tesseral initial perturbation presented in Fig.~\ref{fig:hovmoller_tesseral}. The red dashed line indicates the theoretical frequency of the mode that was excited by the initial perturbation. In the case of T2L8 the limits of the figure exclude this frequency.}
\label{fig:spectra_tess}
\end{center}
\end{figure*}

To explore deeper the properties of the modes developed after imposing the initial perturbation, we use the library SHTns \citep{schaeffer2013efficient} to construct the power spectra of the radial vorticity, $\omega_r = (\nabla \times {\bf u})_r$ computed for each simulation in a shell located at $r=0.64\Rs$. The results are shown in Figs.~\ref{fig:spectra_sec} and \ref{fig:spectra_tess}, corresponding to representative simulations with sectoral and tesseral initial conditions, respectively. The upper and bottom rows of these figures show unnormalized and normalized spectra, allowing comparison between all the exited and the most energetic modes.   

The power spectra presented in Figs.~\ref{fig:spectra_sec} shows the results for simulations SM2, SM4, SM6, and SM8, from left to right, respectively, with sectoral initial perturbations. On each panel of the upper row we observe that the mode with the higher energy corresponds to the mode of the initial sectoral perturbation. Nevertheless, other Rossby modes also appear with frequencies closely following the theoretical dispersion relation. These modes have both, smaller and larger wave numbers than the initial perturbation, indicating direct and inverse energy cascades. Our results confirm the findings of \cite{dikpati2022simulating} who observed inverse cascade effects in shallow-water simulations of Rossby modes excited at the wave numbers of supergranulation.  Our results suggest that energy is transferred towards all available scales.  This can be better observed by comparing the unnormalized spectra, top panels, with the normalized spectra, bottom panels (normalized spectra) panels. In addition to modes following the dispersion relation for Rossby waves, we also observe modes at low frequency and low wave-number, specially for simulations with initial wave numbers larger than 4, panels (g) and (h).  For these cases there are not significant differences in the spectra when it is computed at different radius.

A similar energy cascading of energy is observed in the simulations starting with tesseral modes. Remarkably, the tesseral modes cascade along the dispersion relation of the sectoral modes,  Fig.~\ref{fig:spectra_tess}(a)-(d). It is clear, that the mode with the largest energy is the one imposed as initial condition (top row). However, in this mode, $m=2$, there are several frequencies with larger energy. These frequencies might be the result of non-linear mode interaction between the modes at the bottom and top layers.  This is more evident in the normalized spectra, panels (e)-(h), where the excited modes seem to cross the $y$-axis in $m=0$, as was reported by \cite{bekki2022theory}. 

\subsection{Inertial modes in simulations with differential rotation }
\label{s:dr}

The simulations presented in this section have a resolution of $(128,64,64)$ grid points, they start with all variables set to zero  and the forcing term of Eq.~\ref{eq:body_force} turned on. After a few relaxation times, $\tau_s$, a mean profile of differential rotation  establishes in the domain.  Since perturbations are allowed on both, the potential temperature and the longitudinal velocity, the resulting DR is not exactly similar to the imposed forcing.  Below we explore the excitation of hydrodynamic instabilities for cases with different relative shear defined as $\Omega_2/\Omega_0$. These instabilities are the source of inertial waves. 

\subsubsection{Taylor-Proudman theorem and the adiabaticity of the convection zone}
\label{s:TP}
The turbulent motions resulting in convection may be the source of inertial modes. Although tracking these modes is not impossible \citep{2024ApJ...966...29B}, it is hard to disentangle their properties from those of convection. Additionally,  
convection evinces turbulent viscosity which is much larger than the molecular viscosity estimated for the convection zone. This may affect the dumping of inertial waves, particularly those with small spatial scales. Thus, to better capture the evolution of inertial modes, in the simulations presented below we avoid turbulent convection in the upper layer of the simulation's domain. To do so, we consider for this layer a polytropic index $m\ge1.5$. 

\begin{figure*}
\begin{center}
    \includegraphics[width=\columnwidth]{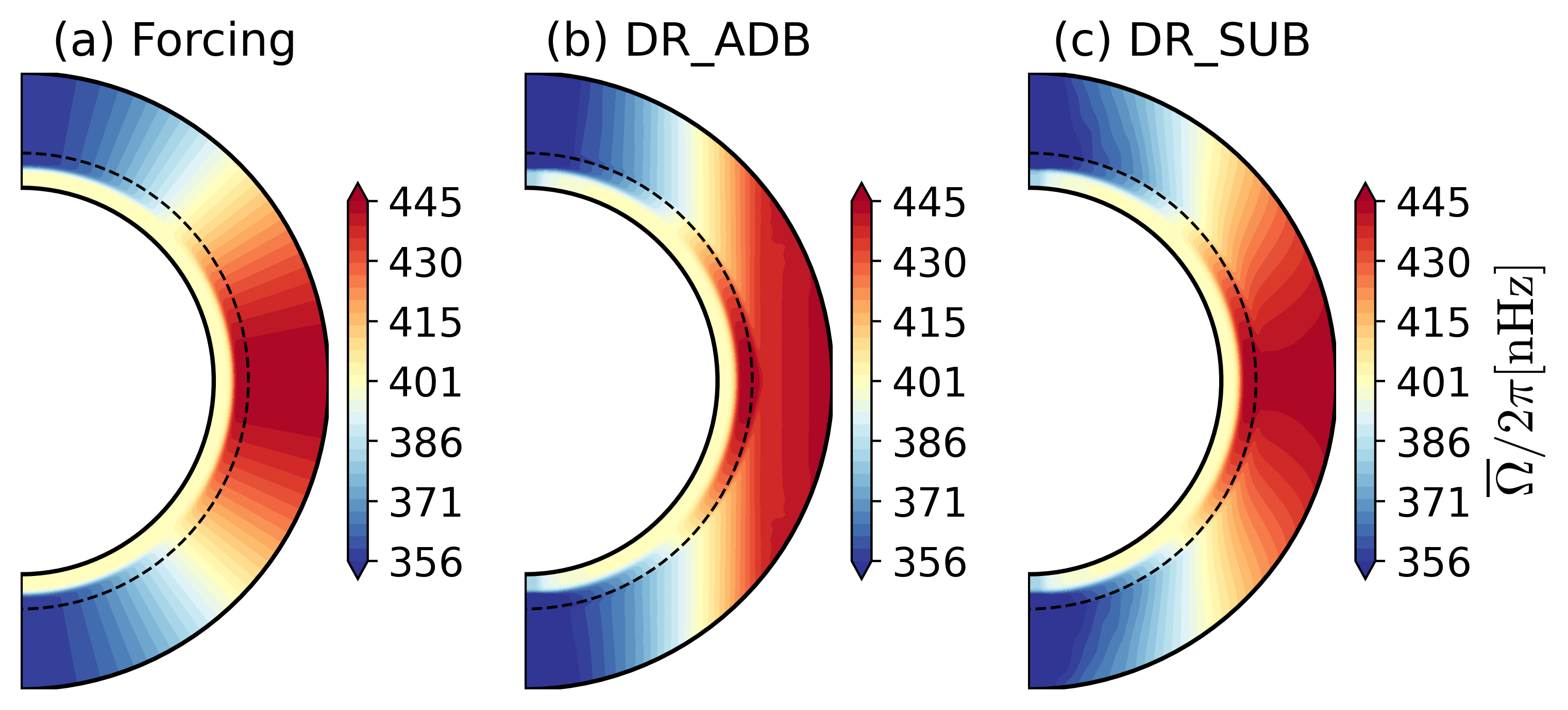} 
\caption{Differential rotation profile in the meridional plane. Panel (a) corresponds to the imposed forcing profile with 20\% of relative shear, Eq.~(\ref{eq:body_force}).  Panels (b) and (c) correspond to simulations where the upper layer is strictly adiabatic and slightly sub-adiabatic, respectively.}
\label{fig:tp_balance}
\end{center}
\end{figure*}

In Fig.~\ref{fig:tp_balance} we present the forced DR profile for a 20\% of relative shear (panel a), and the resulting profiles for simulations where $m_1=1.5$ (b) and  $m_1=1.50001$ (c). In this way, as mentioned above, our forcing profile, Eq.~(\ref{eq:body_force}), together with the thermodynamic stratification defined by Eqs.~(\ref{eq:density_amb}) and (\ref{eq:temp_amb}), generate three different radial layers. The bottom layer of the domain has solid body rotation and highly sub-adiabatic ambient temperature gradient ($m_0=2.5$). The middle layer is also highly sub-adiabatic, yet its rotation is differential. Finally, the upper layer has an ambient temperature gradient either adiabatic, $m_1=1.5$ (panel b) or slightly sub-adiabatic, $m_1=1.5001$ (panel c), and has differential rotation. The tachocline is located in the bottom layer. We noticed that the mean velocity reaches a steady state with the profiles presented in Fig.~\ref{fig:profs_dr} in less than 5 years.  In this stage, the four forces present in the model, namely Coriolis, gravity, pressure gradient and the body force, must balance. In the vertical direction, gravity balances the radial gradient of pressure. In the horizontal direction, the pressure gradient force and the forcing term balance the Coriolis force.  Thus, in order to conserve angular momentum, a meridional circulation (MC) develops together with the DR. In addition,  there are perturbations of potential temperature with a meridional profile satisfying some sort of thermal wind balance,

\begin{equation}
\label{eq:tpbal}
    2 \Omega_0 \frac{\partial \left \langle u_\phi \right \rangle }{\partial z} =   \frac{g}{r\Theta_0}\frac{\partial \left \langle \Theta^{\prime} \right \rangle}{\partial \theta } = 0  \; .
\end{equation}

For the polytropic index $m_1=1.5$, corresponding to a strictly adiabatic uppermost layer, the resulting DR profile presents contours that are aligned with the rotation axis.  This means that the upper layer of the domain obeys the Taylor-Proudman theorem. In this case both sides of Eq.~\ref{eq:tpbal}, namely the inertial and the baroclinic terms, vanish.

Conversely, if $m_1$ is slightly larger than $1.5$, a certain amount of baroclinicity is observed. Thus, the Taylor-Proudman theorem is broken and the DR profile develops contours tilted with respect to the rotation axis. At lower (higher) latitudes the contours are tilted away (towards) the rotation axis.  In the Sun, all contours are tilted away from the axis as shown in Fig.~\ref{fig:tp_balance}(a).  In additional simulations (not shown here), we found that larger values of $\tau$ reverse the tilt of the $\overline{\Omega}$ contours at higher latitudes, causing them to point away from the rotation axis. This modification, however, does not affect the evolution or the spectral properties of the inertial modes discussed below, apart from their amplitude. We also verified that the results are only weakly sensitive to variations in $\tau_s$. For this reason, we defer a detailed parametric analysis of these variables to a companion paper, where the forcing is prescribed by the observed differential rotation (Finotti et al., in prep.).

Recent numerical findings suggest that the lower portion of the solar convection zone may be convectively stable according to the Schwarzschild criterion, yet capable of transporting heat efficiently due to the downward enthalpy flux driven by overshooting plumes \citep{2017ApJ...845L..23K}. Furthermore, a subadiabatic stratification at the base of the convection zone has been proposed as a condition for the excitation of the polar inertial mode with azimuthal order m=1, which manifests as a large-scale vortex structure \citep{gizon2021solar, 2021ApJ...908..160H}. Motivated by these considerations, and by the fact that a slightly subadiabatic profile can induce tilted rotational contours, the ambient stratification adopted in the simulations presented below includes a mildly subadiabatic thermal stratification in the upper layer of the domain. The preferential development of baroclinicity in subadiabatic layers might be due to the fact that gradients of potential temperature developed by thermal wind balance may persist in the layer even in the presence of waves or perturbations. Conversely, in strictly adiabatic layers, marginally stable to convection, even small perturbations may quickly homogenize the gradients of potential temperature. The same is true for super-adiabatic convective layers.  
The sustainment of potential temperature gradients at higher latitudes may of course allow for potential energy availability when parcels of fluid displace in the proper direction. When this energy is transferred to the fluid motions, it develops the so-called baroclinic instability.

\subsubsection{Baroclinic instabilities}
\begin{figure}
\begin{center}
	\includegraphics[width=0.6\columnwidth]{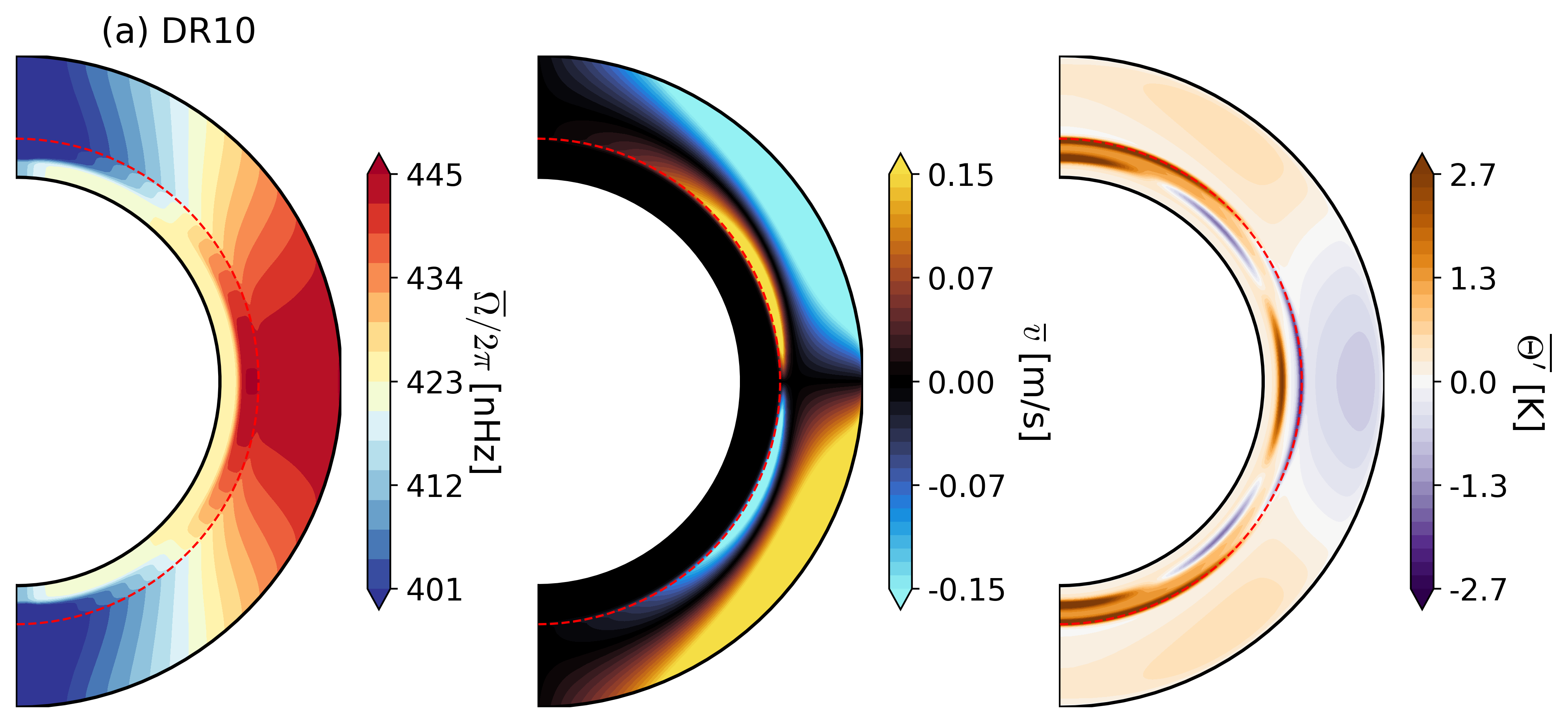}\\
    \includegraphics[width=0.6\columnwidth]{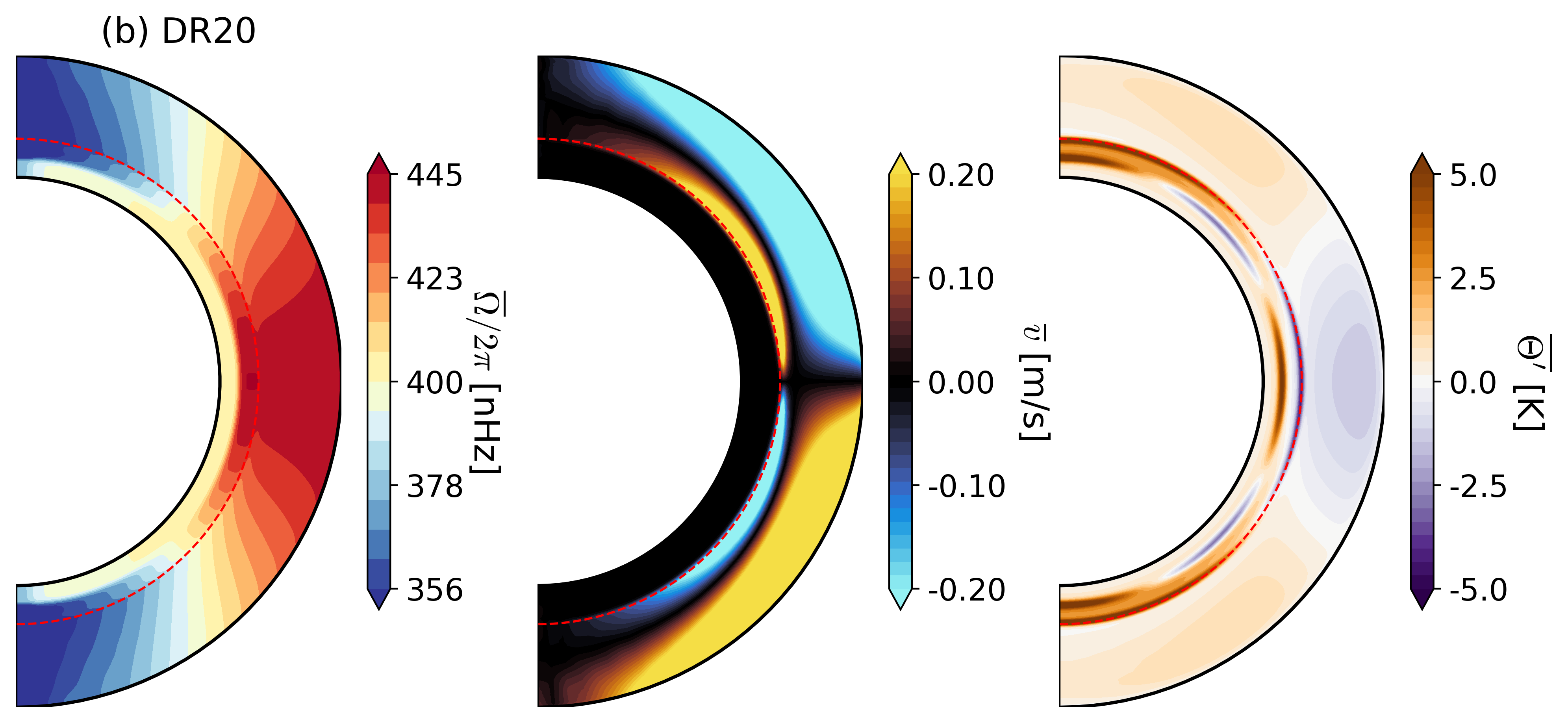}\\
    \includegraphics[width=0.6\columnwidth]{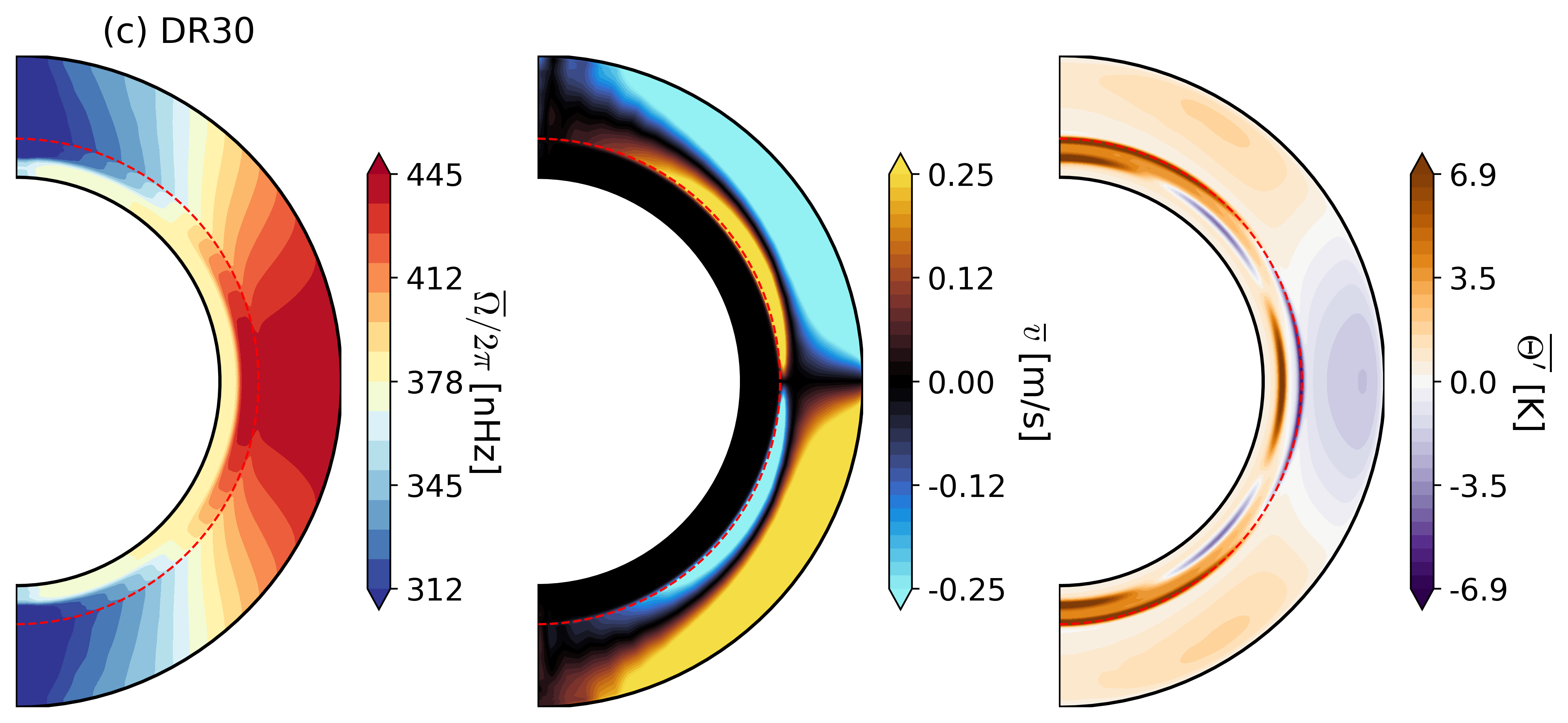}\\
    \includegraphics[width=0.6\columnwidth]{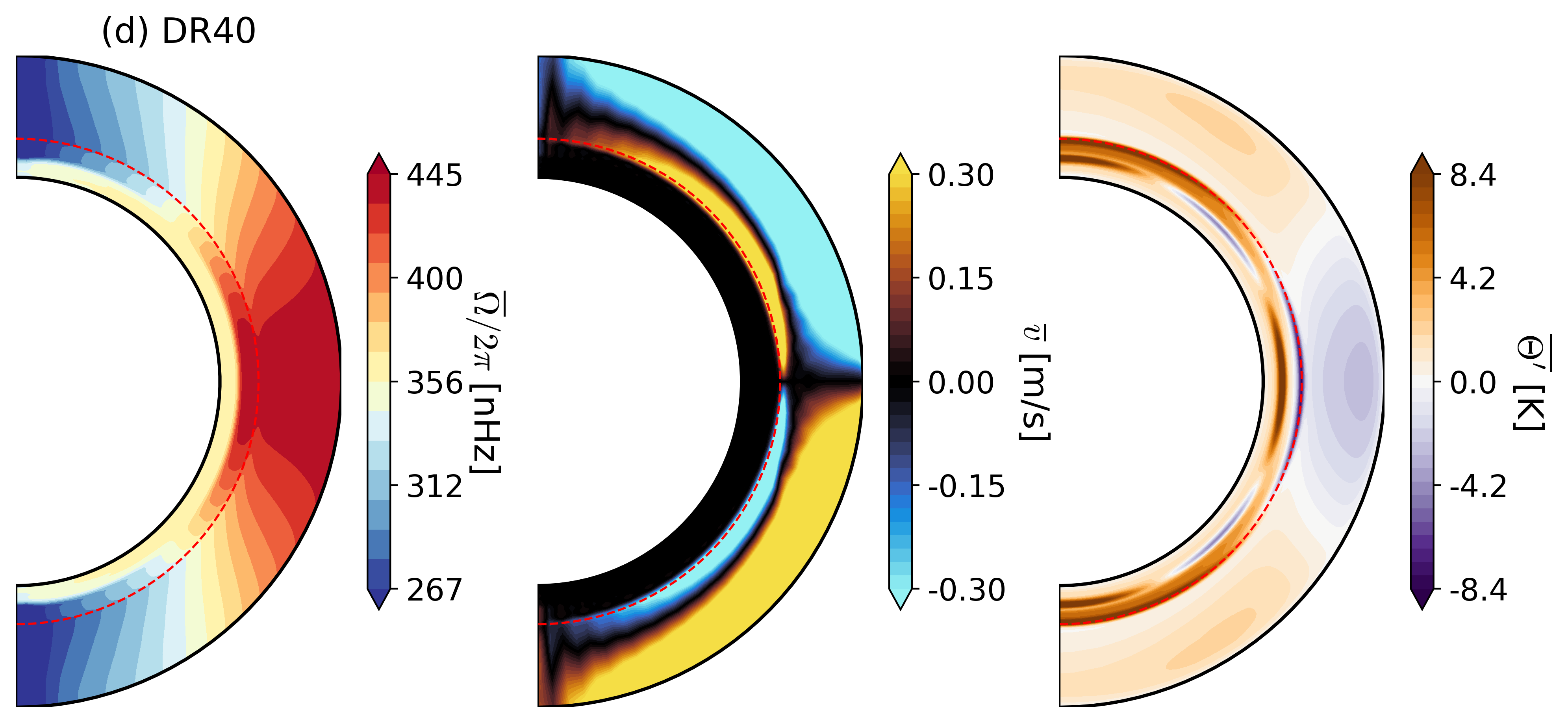}\\
    \caption{Meridional profiles of DR (left side of each panel) and MC (middle) and $\overline{\Theta^{\prime}}$ (right side) for characteristic simulations with different shear  (10\% (a) - 40\% (d)).  Note that the color table has different limits for each panel.}
    \label{fig:profs_dr}
\end{center}
\end{figure}

To explore the role of shear in the development of inertial modes, we consider simulations with relative shear between 5\% and 40\%, which are called DR05, DR10, ..., DR40.  The resulting profiles of DR profile, $\overline{\Omega}/2\pi$, in the meridional plane for some characteristic simulations is presented in the left panels of Fig.~\ref{fig:profs_dr} (a) to (d). Here and in Fig.~\ref{fig:profs_dr} the overline represents average over longitude and time during the steady state of the simulations. Note that there is not apparent difference in the meridional DR profile, yet the scale of the contours is different on each plot (see color table). Evidently, increasing the latitudinal shear also produces an increasing radial shear at tachocline levels.  The middle and right panels in Fig.~\ref{fig:profs_dr} show the developed meridional circulation, represented by the longitudinal velocity, $\overline{v}$, and the distribution of potential temperature perturbations, $\overline{\Theta^{\prime}}$, in the meridional plane.  Both quantities increase for stronger shear. In all cases the meridional flow shows one cell per hemisphere with anticlockwise circulation, the cell extents closer to the poles when the shear is stronger, however, even for the stronger shear, the amplitude of $\overline{v}$ reaches only $\sim 0.4$ m s$^{-1}$. As it will be shown below, this weak MC is sufficient to balance the external body force which pumps angular momentum towards the equator.  Similarly, in order to balance the radial variations of the angular velocity and ensuring thermal wind balance, a latitudinal gradient of $\overline{\Theta^{\prime}}$ also develops. These gradients are stronger at tachocline levels and weaker at surface levels. Similarly, only few K are necessary to balance the strong tachocline radial shear.  

\begin{figure}
\begin{center}
	\includegraphics[width=0.8\columnwidth]{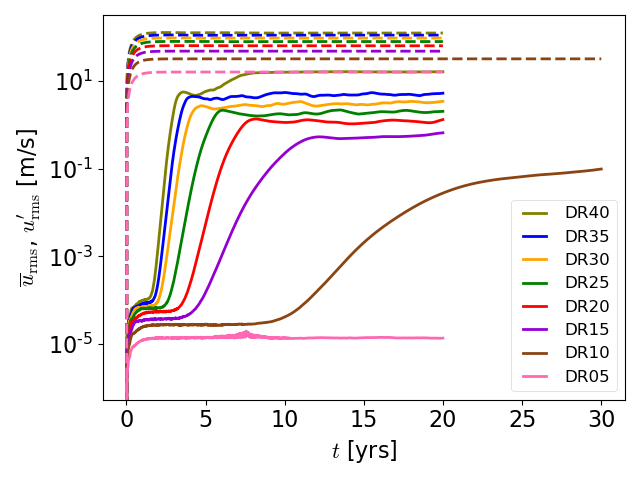}
    \caption{The dashed lines represents $U_{\rm rms}$ and lines represents $U^\prime_{\rm rms}$ }
    \label{fig:mnturbvel}
\end{center}
\end{figure}

Figure~\ref{fig:mnturbvel} shows the time evolution of the ${U}_{\rm rms}=\sqrt{\brac{u^2 + v^2 + w^2}_{\phi\theta}}$, and $U^{\prime}_{\rm rms}=\sqrt{\brac{u^{\prime 2} + v^{\prime 2} + w^{\prime 2}}_{\phi\theta}}$, where the prime indicates the turbulent part of a quantity, e.g.,  $u^{\prime} = u - \mean{u}$ (this encompasses all the modes different from $m=0$).  The angular brackets $\brac{}_{\phi\theta}$ correspond to longitudinal and latitudinal average.

\begin{figure*}
\begin{center}
	\includegraphics[width=0.3\columnwidth]{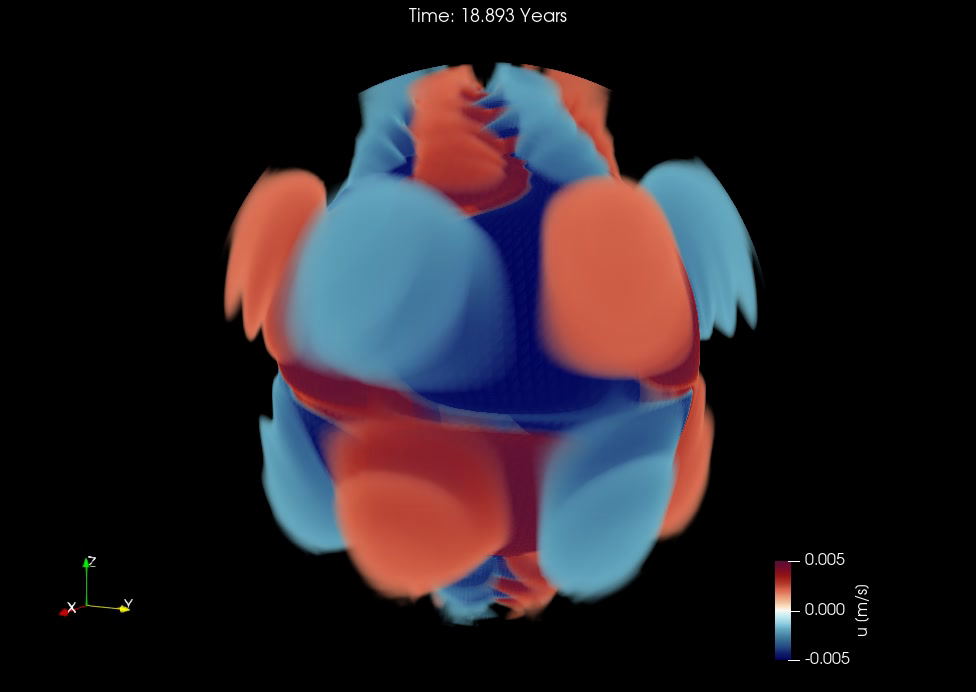} 
	\includegraphics[width=0.3\columnwidth]{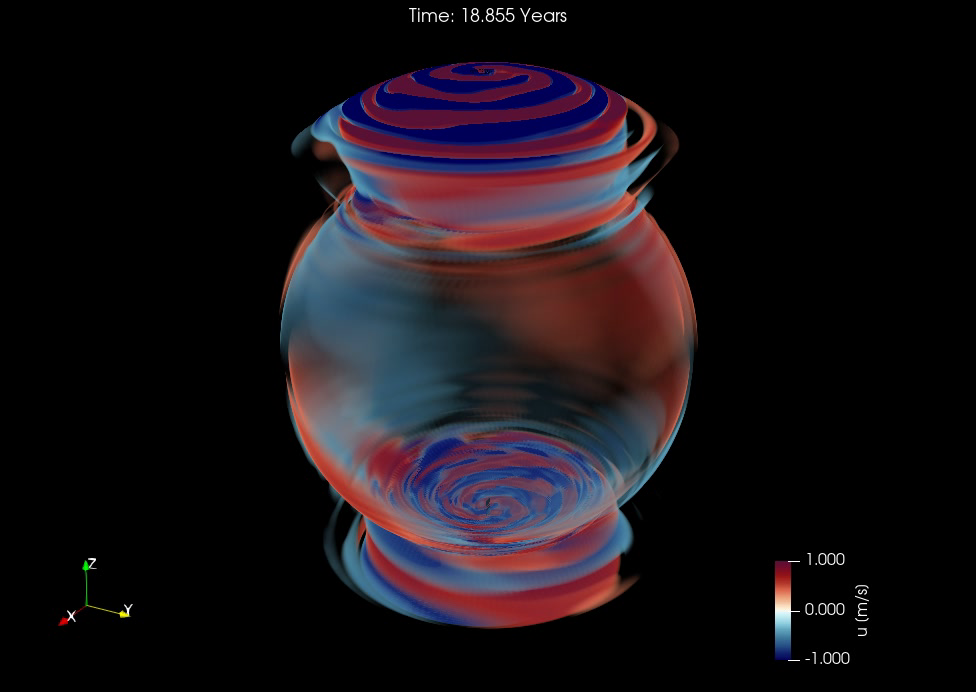}
	\includegraphics[width=0.3\columnwidth]{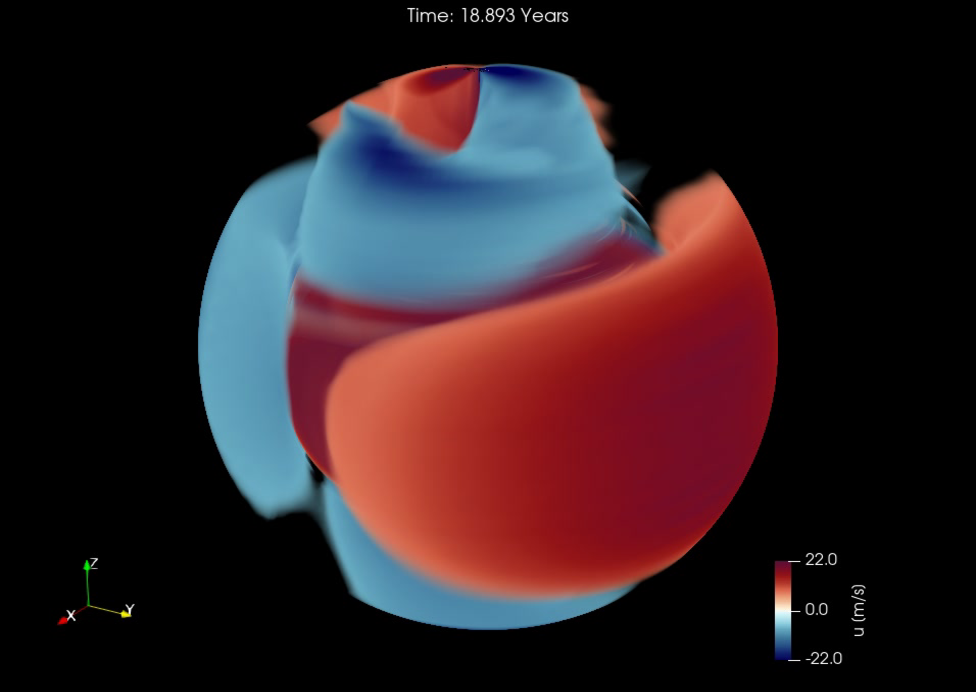} 
    \caption{Snapshots of the longitudinal velocity $u^{\prime}$ for characteristic simulations DR10, DR20 and DR40, from left to right, respectively.  The snapshot is captured during statistically steady state, at the end of the simulation. See animations corresponding to these simulations in \url{ https://zenodo.org/records/18388135}. }
    \label{fig:snapshots}
\end{center}
\end{figure*}

After a few years, $U^{\prime}_{\rm rms}$ starts to increase due to instabilities in the shear. The growth rate of instabilities increases with the amount of shear imposed to the system, as can be noticed in Fig.~\ref{fig:mnturbvel}.  The figure shows that the mean quantities (continuous lines) reach steady state in less than 5 years for all simulations. This is expected after the considered value of $\tau_s$.  Except for simulation DR05, the turbulent velocities start growing in amplitude and reach a saturation value which is larger for simulations with stronger shear.  The snapshots presented in Fig.~\ref{fig:snapshots}, for simulations DR10, DR20 and DR40, from left to right, respectively,  represent the longitudinal velocity, $u^{\prime}$ which depicts the modes persistent in the steady state of the simulations. 

The animations corresponding to the simulations presented in Fig.~\ref{fig:snapshots} are presented in \url{https://zenodo.org/records/18388135} are the best way to follow the development of these unstable modes.  They start appearing at the base of the convection zone where the radial shear is stronger and at polar latitudes.  The growing perturbations appear in the form of polar vortices that have clockwise orientation in the northern hemisphere.  These modes are retrograde, i.e., in longitude, they propagate in the opposite direction of $\Omega_0$. The animations show that, in some cases, they also propagate radially, by crossing the entire uppermost layer and also penetrating into the underneath stable layer. Nevertheless, no radial modes are observed.  

The radial propagation of vortex motions is faster in simulations with stronger imposed shear. At lower latitudes, the dominant modes are predominantly symmetric with respect to the equator: the radial velocity perturbation \( u' \) changes sign across the equator, while the azimuthal component \( v' \) retains the same sign. These low-latitude modes extend throughout the convection zone only under conditions of weak shear. In simulation DR20, for instance, high-latitude modes dominate, and low-latitude modes appear only near the tachocline.

This pattern shifts in simulations DR35 and DR40, where the amplitudes of high- and low-latitude modes become comparable. In DR35, polar vortices exhibit multiple longitudinal wavenumbers, while at low latitudes the flow exhibits temporal variability in equatorial symmetry, alternating between symmetric and antisymmetric structures. Simulation DR40 is clearly dominated by a longitudinal mode with azimuthal wavenumber \( m = 1 \) across all latitudes and displays a three-band latitudinal structure. These results suggest the occurrence of a bifurcation in the instability as the imposed shear increases from 30\% to 40\%, leading to a change in the symmetry properties of the modes. Simulation DR35 likely represents a transitional regime, consistent with its observed symmetry migration.

\begin{figure}
\begin{center}
	\includegraphics[width=0.8\columnwidth]{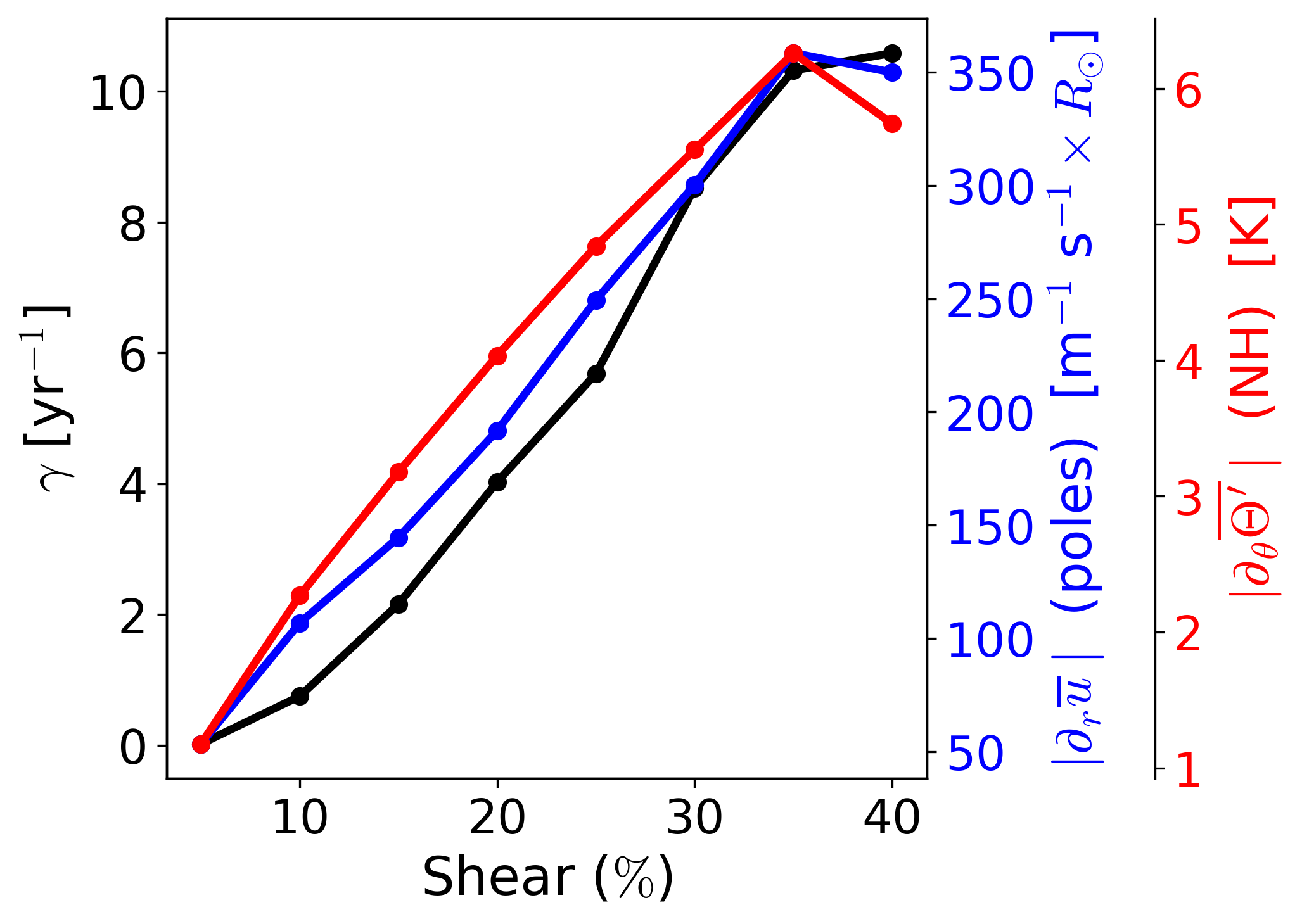}
    \caption{Black line: growth rate of the unstable modes, $\gamma$, computed from the time evolution of ${\bf u}^{\prime}_{\rm rms}$ as a function of the shear parameter. Blue and red lines lines:  absolute value of the radial derivative of $\overline{u}$ at polar latitudes and  absolute value of the latitudinal derivative of $\theta^{\prime}$ at polar latitudes. Both quantities are plotted as a function of the shear.}
    \label{fig:wrate}
\end{center}
\end{figure}

The growth rate of the instability, $\gamma$, is computed from the time evolution of ${\bf u}^{\prime}_{\rm rms}$ (see Fig.~\ref{fig:mnturbvel}) and is shown as a function of the imposed shear in Fig.~\ref{fig:wrate} (black line). It is evident that the threshold for instability lies between 5\% and 10\%, and that $\gamma$ appears to saturate for shear values exceeding 35\%. 

Since the unstable modes first emerge at the poles—where the radial shear is the strongest—we examine whether the growth rate is consistent with that of a baroclinic instability. To do so, we compare the trend of $\gamma$ with that of the radial shear at the poles, represented by the blue line in Fig.~\ref{fig:wrate}. For a system in thermal wind balance, the radial shear is expected to be proportional to the latitudinal gradient of potential temperature, as described by Eq.~\ref{eq:tpbal}.

The red line in Fig.~\ref{fig:wrate} shows $\partial \Theta^{\prime}/\partial \theta$ evaluated over the polar region ($0^{\circ} \le \theta \le 5^{\circ}$). There is a clear agreement between the blue and red curves, both of which also show a strong correlation with the behavior of $\gamma$. The saturation of the growth rate for shear values above 35\% is apparent in all trends.
As we will demonstrate later, the saturation may be due to the development of larger Reynolds stresses, which react back the imposed angular velocity by pumping angular momentum towards the poles and therefore diminishing the radial shear.

\subsubsection{Morphology of the polar vortex}

Polar vortices appear to be ubiquitous in rotating fluid spheres. They have been observed in Earth’s atmosphere, in the gaseous planets, and in the Sun. If their generation arises from baroclinic instability, their presence is expected: in Earth’s atmosphere and in the gas giants due to differential solar heating. In the Sun there is no comprehensive explanation for the existence of a latitudinal gradient of potential temperature (viz. entropy) in the convection zone. Nevertheless, the tilted contours of differential rotation suggests its existence. Underneath the convection zone, at tachocline levels, if the Sun is in thermal wind balance, the latitudinal gradient of $\Theta$ must also be present. Therefore, the polar vortices should also be expected.  

Of particular relevance for understanding inertial modes is the morphology of these vortices. In the solar case, observational evidence \citep{hathaway2013giant,gizon2021solar} indicates that in the northern hemisphere the vortex has a counterclockwise orientation, with the largest energy contained in modes $m=1$, $2$, and $3$, at frequencies of $-85$, $\sim -151$, and $\sim -224$ nHz in the Carrington frame, respectively. These structures have been detected as deep as $0.6 \, R_{\odot}$ \citep{2024ApJ...967...46M}, and their amplitudes vary with the solar cycle \citep{2025ApJ...989...26D}.  

In our simulation with 20\% shear—an analogue to the Sun—at the equator, where the rotation is $\Omega_0$, the frequencies of the first three longitudinal modes are $-78.3$, $\sim -125.9$, and $\sim -164$ nHz, in fair agreement with observations. However, Fig.~\ref{fig:snapshots} (middle panel) and its corresponding animation show that the orientation of the vortex is clockwise.  

\begin{figure*}
\begin{center}
	\includegraphics[width=0.36\columnwidth]{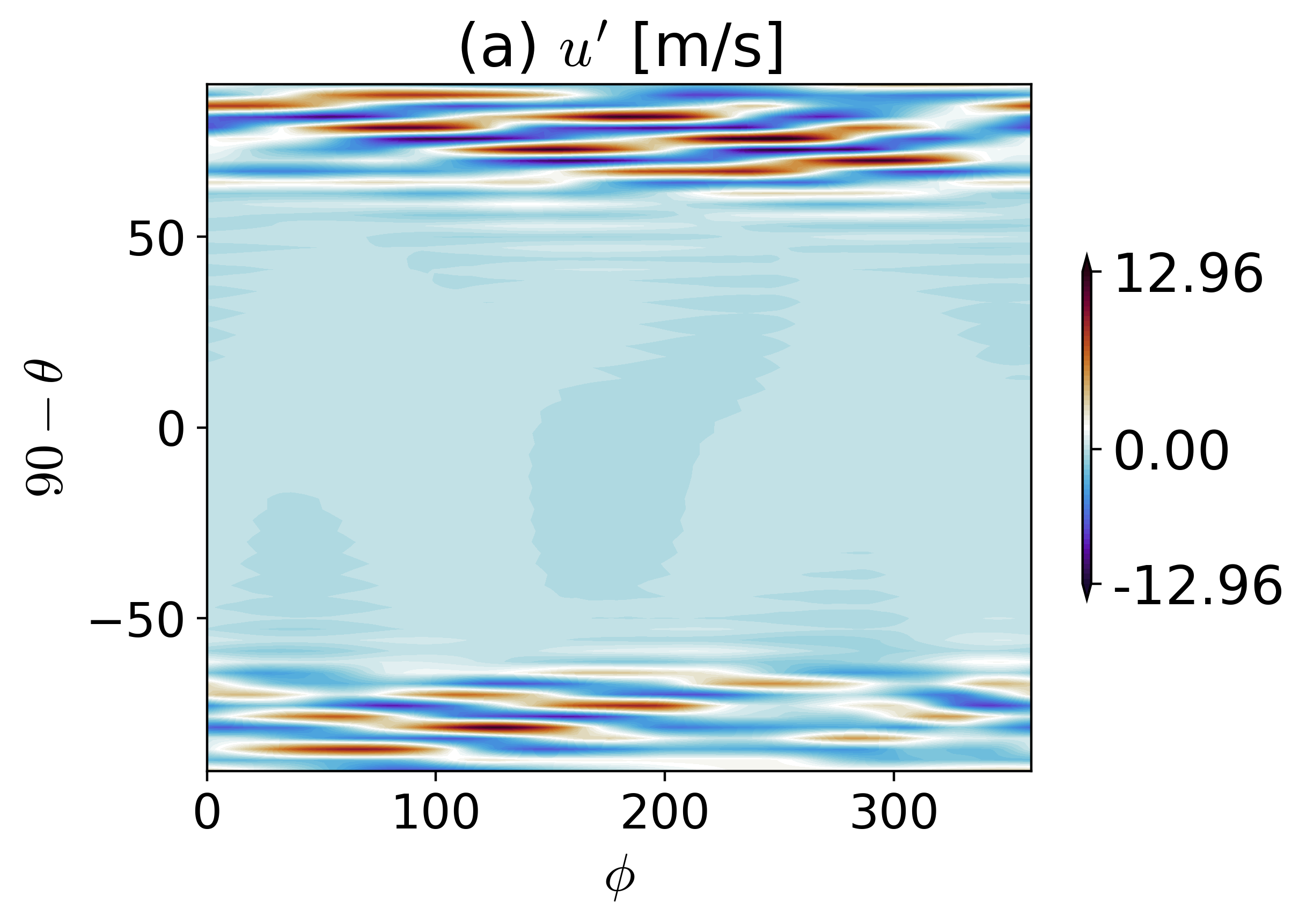} 
	\includegraphics[width=0.3\columnwidth]{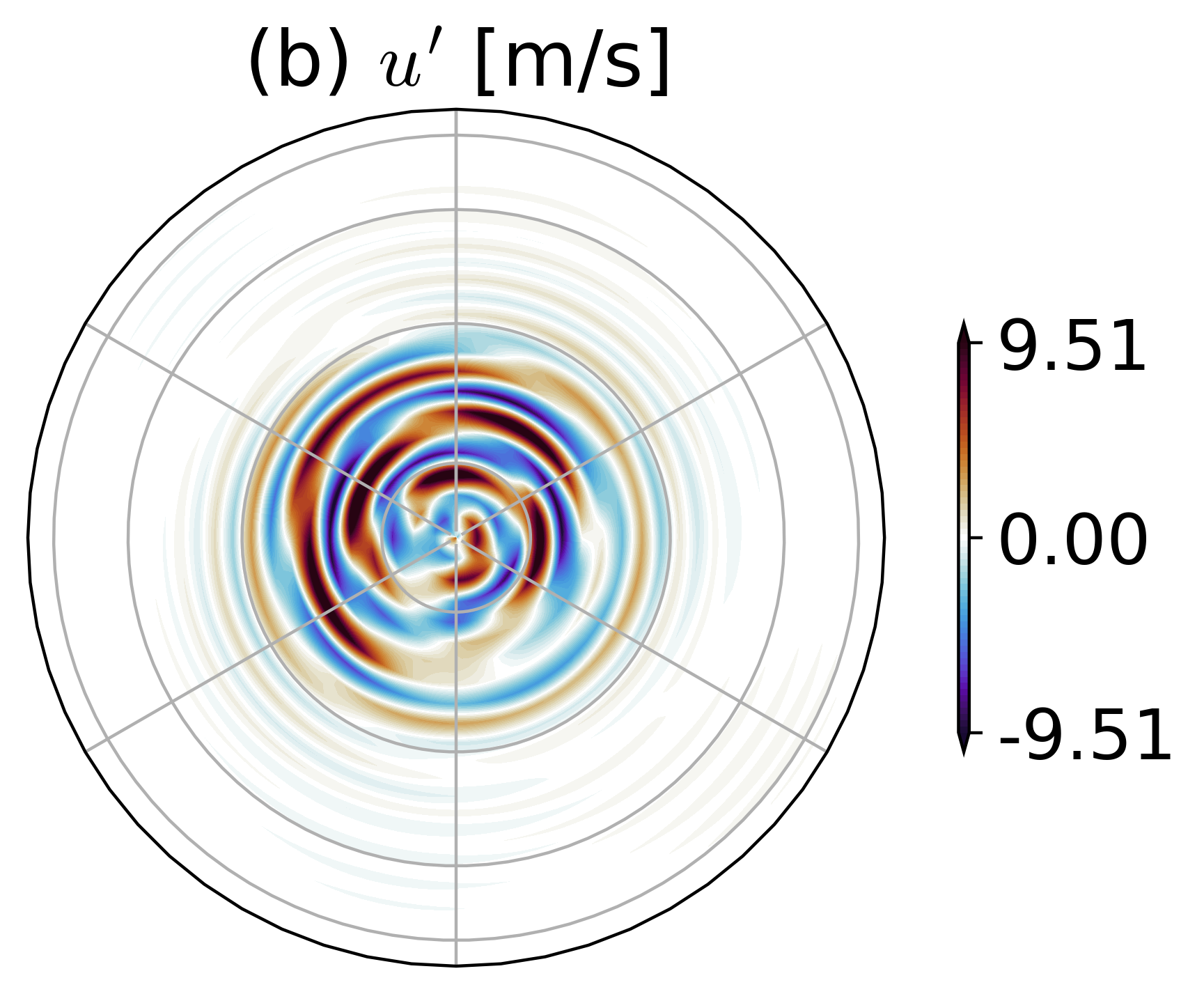}
	\includegraphics[width=0.3\columnwidth]{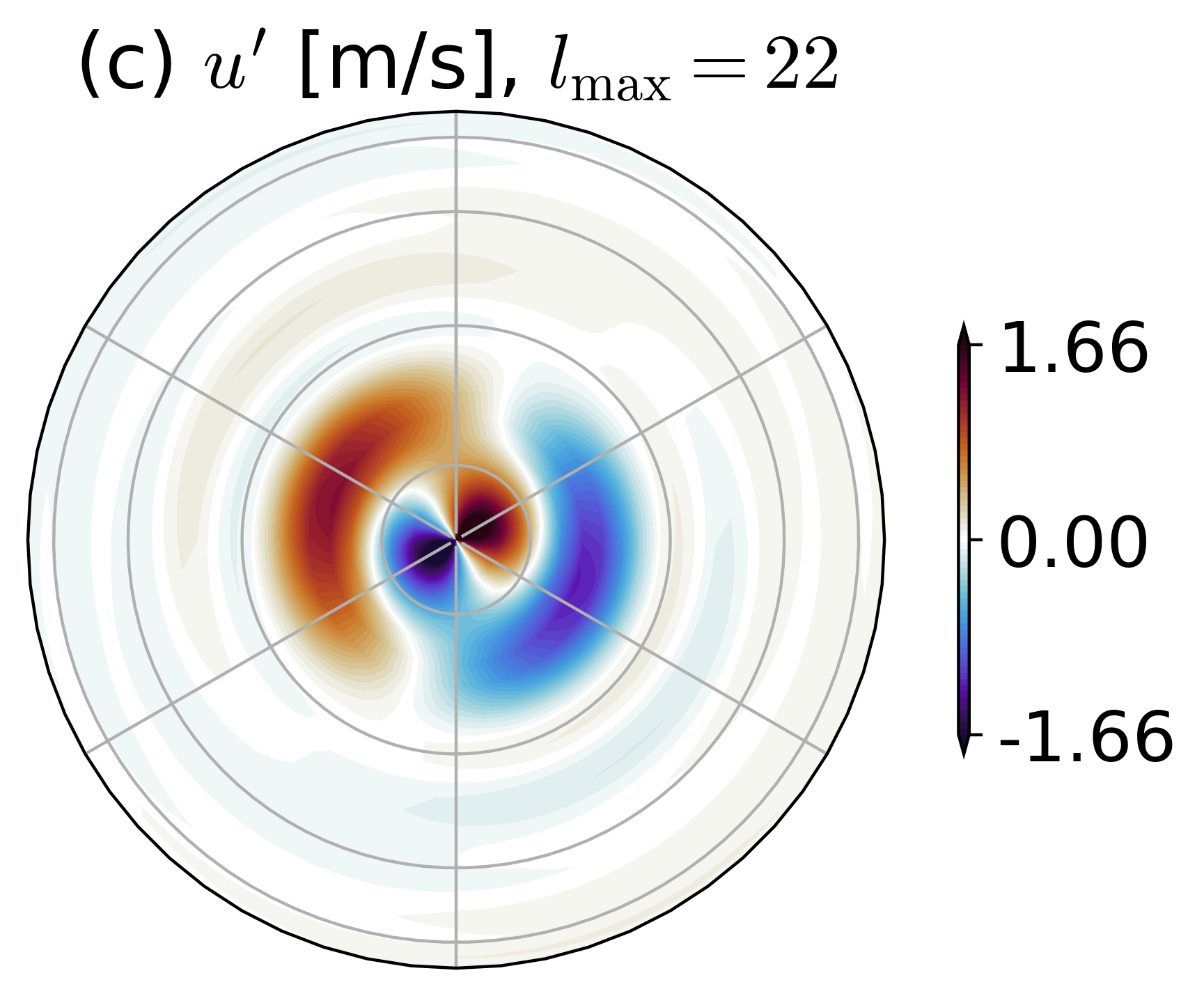} 
    \caption{Snapshot of longitudinal velocity $u^{\prime}$ for simulation DR20 in the ($\phi,\theta$) plane (a), and in orthographic projection (b) and (c). In panel (c) the velocity field is decomposed into spherical harmonics, with only the $m=1$ mode and angular degree up to $l=22$. The snapshot corresponds to $t \sim 19$ years of evolution.}
    \label{fig:vortex}
\end{center}
\end{figure*}

A simple visualization of the solar vortex orientation is provided in Fig.~5(a) of \cite{2025ApJ...989...26D}. In a latitude–longitude plane, a counterclockwise vortex appears as fluid stripes extending from west to east, connecting high latitudes ($\sim \pm 70^{\circ}$) with mid-latitudes ($\sim \pm 50^{\circ}$). We reproduce this diagram in Fig.~\ref{fig:vortex}(a). In our case,  the stripes extend from intermediate to high latitudes. Panel (b) shows the vortex at the north pole in an orthographic projection, where its orientation is clearly clockwise. 

It is worth noting that the stripes in panel (a) and the vortex arms in panel (b) display pulses of enhanced velocity magnitude that propagate equatorward. To isolate the evolution of these more intense modes, we filtered the data using spherical harmonics. Considering modes with $m=1$ and different latitudinal orders, we find that the pulse direction is counterclockwise up to $l=28$. Figure~\ref{fig:vortex}(c) shows the filtered velocity field, for $m=1$ with $l_{\rm max}=22$, where the vortex orientation is consistent with observations. We emphasize, however, that these filtered modes contain only about $20\%$ of the total energy present in the unfiltered polar vortex.

\cite{gizon2021solar} and \cite{bekki2024sun}, based on hydrodynamic simulations that incorporate estimates of turbulent viscosity in the solar convection zone and the tachocline, argue that both, the orientation and amplitude of the modes, depend on the degree of subadiabaticity at the base of the convection zone. \cite{bekki2024sun} further suggest that at the tachocline, $\delta$ (see definition in \S\ref{s:model}) must lie between $-2\times 10^{-5}$ and $-2.4\times 10^{-5}$ in order to reproduce the observed mode frequencies and amplitudes. However, it is important to note that recent helioseismic inversions indicate that the radial shear layer driving the baroclinic instability is located not at the base of the convection zone, but within the radiative zone. In that region, $\delta$ is expected to differ significantly from the proposed values. Moreover, given that convective velocities are much smaller than those predicted by mixing-length theory, the value of the turbulent viscosity is difficult to constrain at such depths \citep{2024PhFl...36k7136B,2025arXiv250505454S}. Most importantly, this strongly sheared region is also expected to host a substantial magnetic field, which interacts with the fluid motions and modifies the properties of the polar vortices \citep{2024PNAS..12115157D}.  

In our model, $\delta = 1\times10^{-5}$ in the uppermost, slightly subadiabatic, layer (not far from the values suggested by \cite{bekki2024sun}), while at the location of the shear layer it reaches $-1 \times 10^{-1}$, consistent with a subadiabaticity of the Sun's radiative zone. Our findings show that in a well-constrained model, the observed modes are indeed present but remain hidden within more energetic higher-order modes. This suggests that some ingredients may be still missing from our setup. For example, the effective viscosity in our simulation might not match the values expected in the upper radiative zone. Our results may also reflect the lack of magnetic influences. To avoid extending the scope of this paper, we leave this issue for future works. Explanations based on finely tuned solutions would be premature given the current state of observational and theoretical understanding of the solar interior and the behavior of convection and inertial waves.

\subsubsection{Spectra of the inertial modes}

The development of the baroclinic and perhaps other instabilities results in perturbation flows that are subject to the Coriolis force and propagate as inertial waves.  In this section we explore the frequency of propagation, $\omega$ of different longitudinal modes, $m$. The analysis is more complex in this case than in the case of solid body rotation, \S\ref{s:solid}, since different modes appearing at different latitudes and radius have different frequencies. Note, additionally, that each radius and latitude have a different angular velocity, therefore the modes also suffer Doppler shifting. We noticed however that the spectra of these modes are the same if computed for any of the non-axisymmetric velocity components than if computed for the vertical vorticity, as in Figs.~\ref{fig:spectra_sec} and \ref{fig:spectra_tess}. In the following analysis we use $v^{\prime}$. 
 
\begin{figure*}
\begin{center}
	\includegraphics[width=0.48\columnwidth]{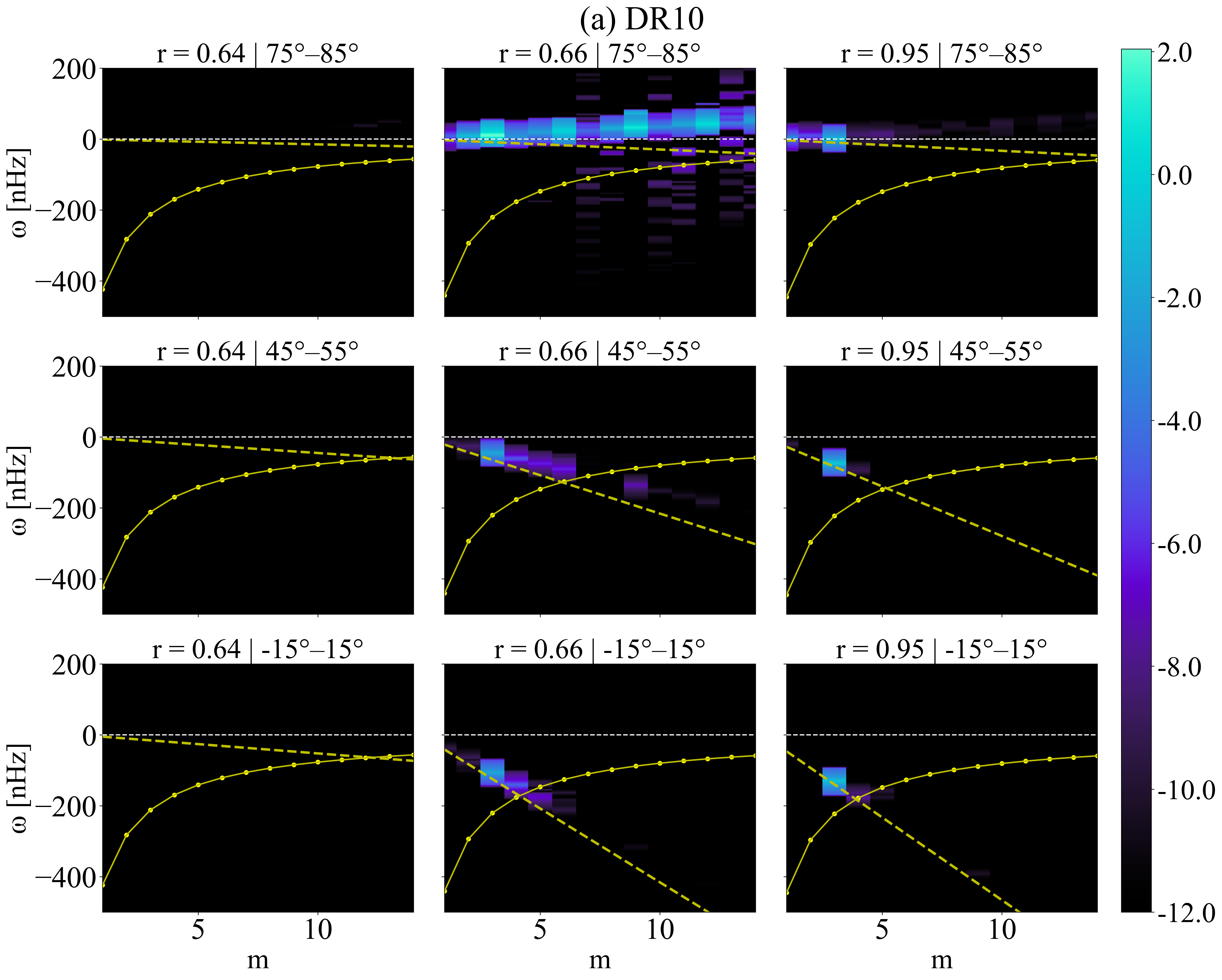} 
	\includegraphics[width=0.48\columnwidth]{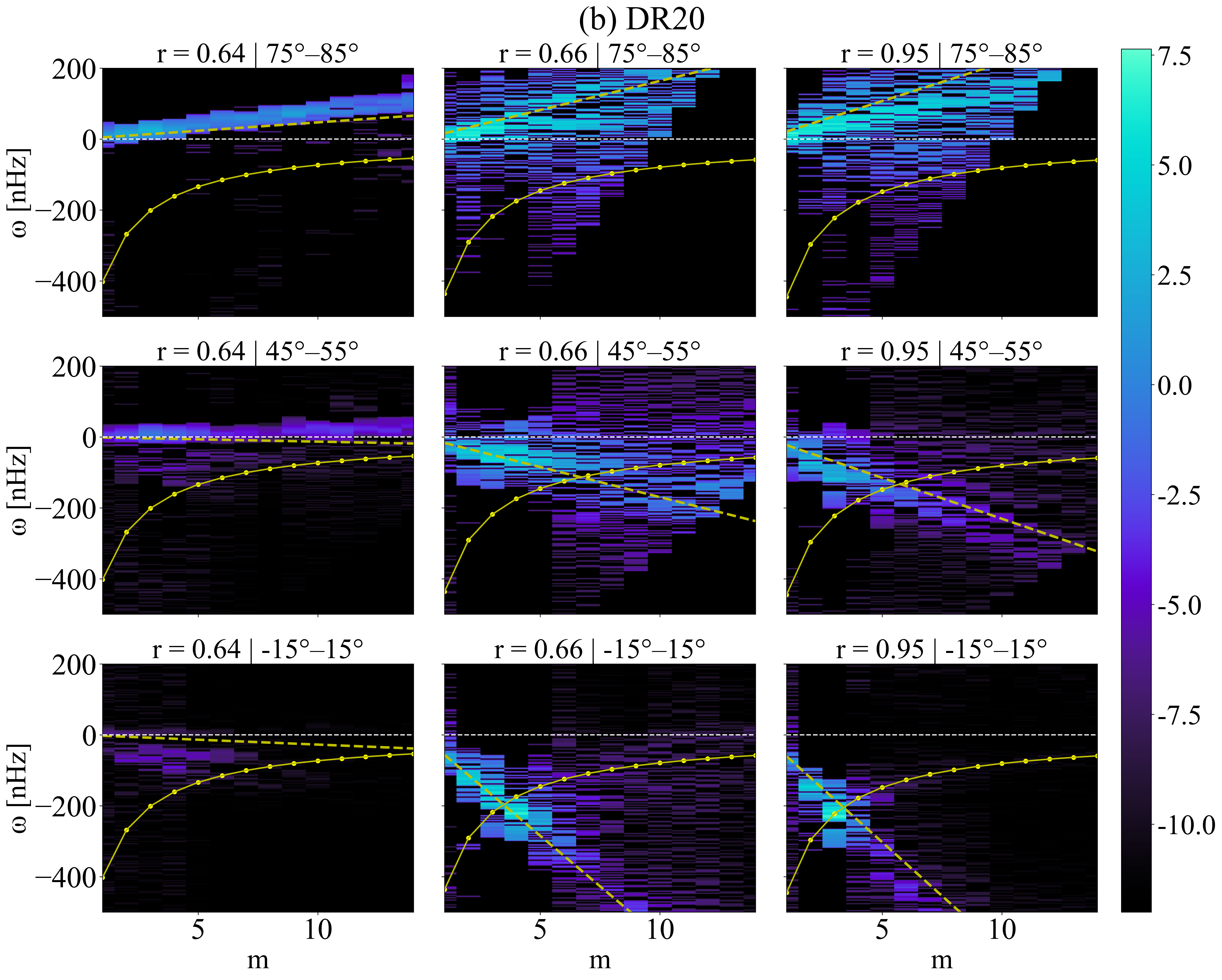}\\
	\includegraphics[width=0.48\columnwidth]{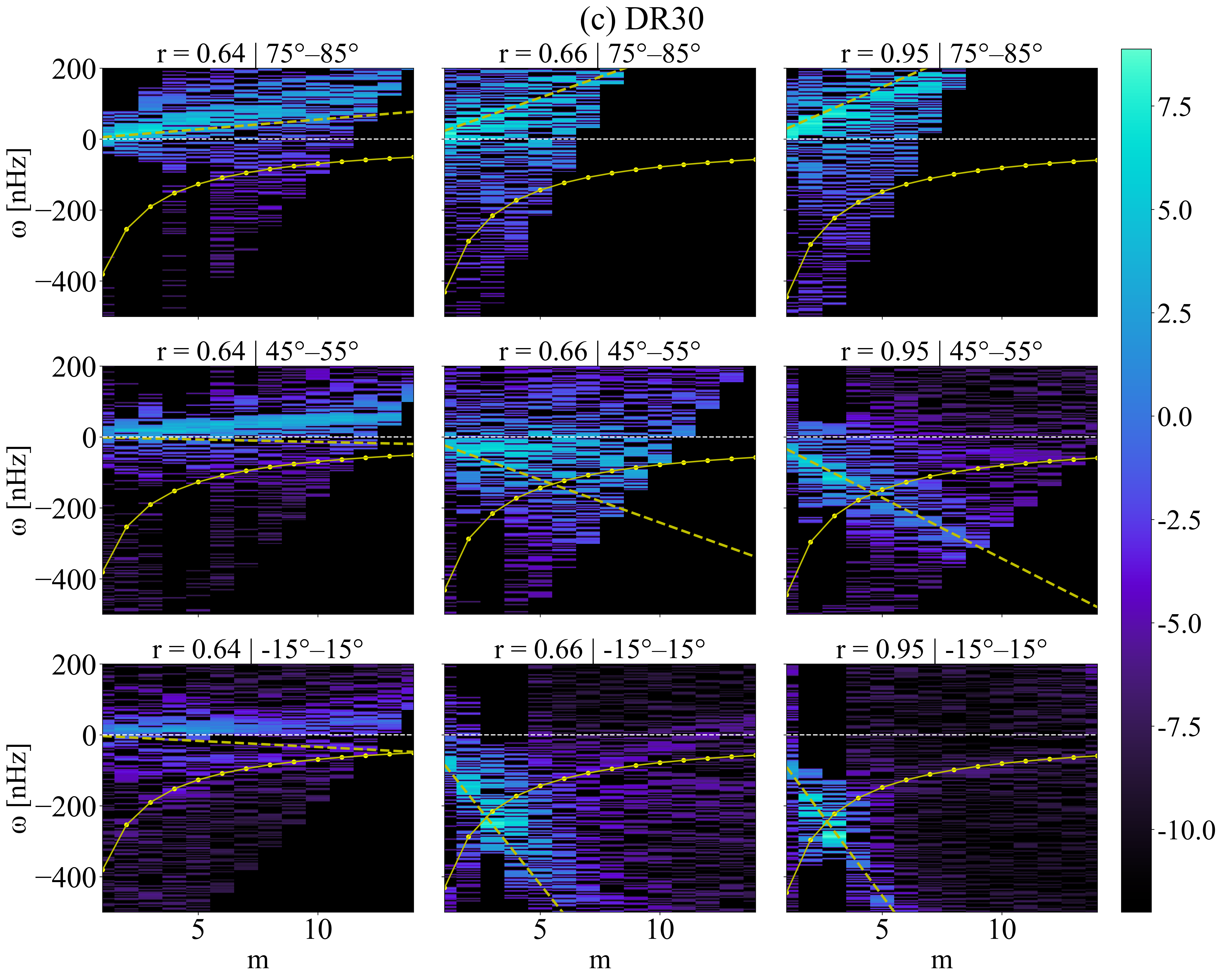} 
	\includegraphics[width=0.48\columnwidth]{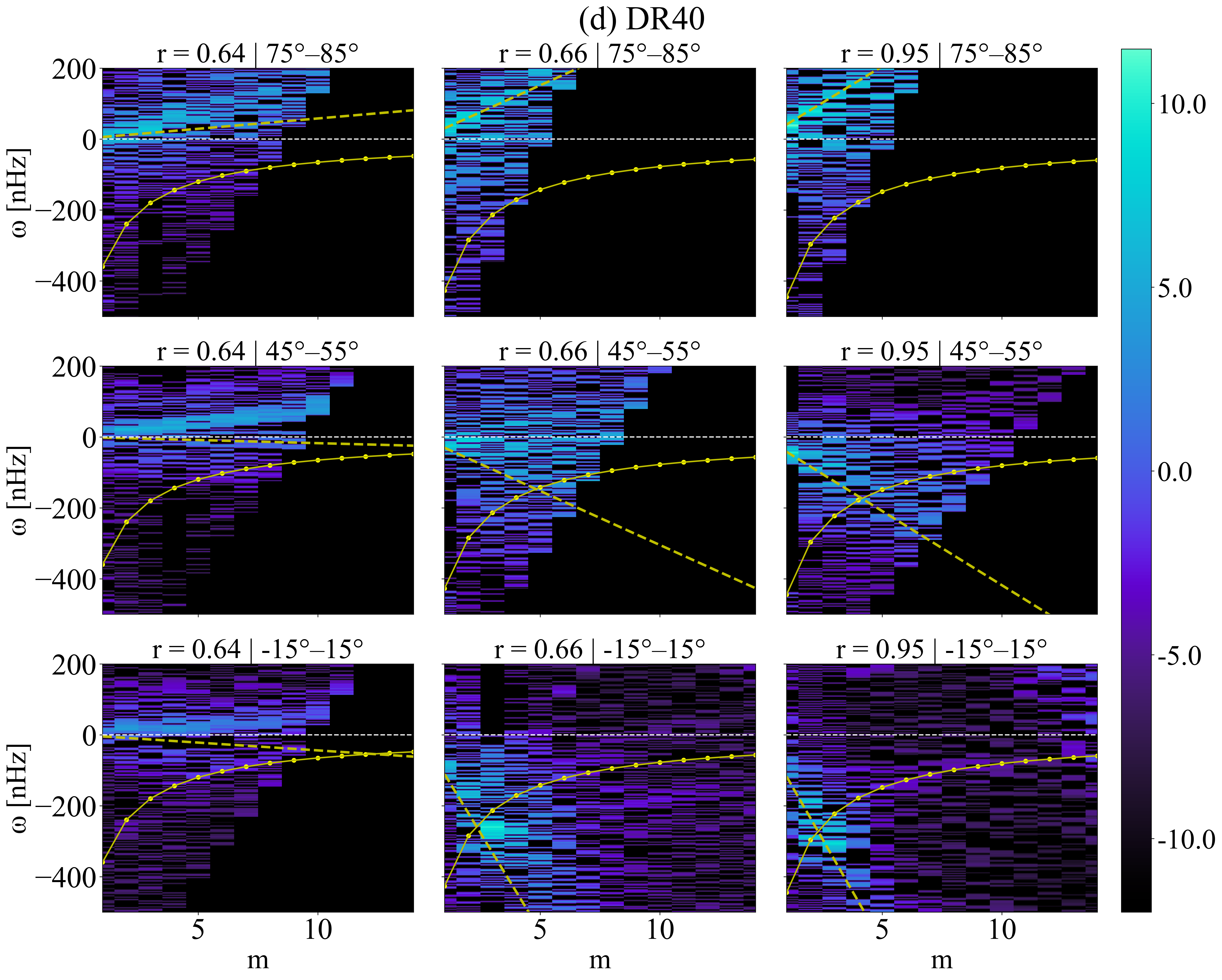} 
    \caption{Power spectra depicting the temporal frequency $\omega$ as a function of the longitudinal wavenumber $m$ for four characteristic simulations, DR10 (panel a), DR20 (b), DR30 (c) and DR40 (d). Each panel displays nine spectra for three different radii, $r/R_\odot = 0.64$, $0.66$, $0.95$, from left to right, and three different latitude bands, 75$^\circ$ to 85$^\circ$ (pole), 45$^\circ$ to 55$^\circ$ (intermediate latitudes), and $-15^\circ$ to 15$^\circ$ (equator), from top to bottom. The yellow lines show the theoretical dispersion relation for equatorial (solid) and high latitude (dashed) modes. Both, computed frequencies and dispersion relations, have been corrected for Doppler shift on each region.}
    \label{fig:sp_shear}
\end{center}
\end{figure*}

To obtain the nine $(m,\omega)$ power spectra displayed in Fig.~\ref{fig:sp_shear}, panels (a)  to (d), we extracted consecutive turbulent snapshots of $v^{\prime}$ from each simulation at the statistically steady state and processed them with a dedicated {\sc python} routine.  For the three target radii discussed in \S\ref{s:model} ($r/R_\odot = 0.64$, 0.66, 0.95) and three latitude bands (75$^\circ$ to 85$^\circ$, 45$^\circ$ to 55$^\circ$, $-15^\circ$ to 15$^\circ$), the code (i) performs an orthonormal discrete Fourier transform in longitude, (ii) retains azimuthal wavenumbers $1 \le m < 15$, and (iii) integrates the complex amplitudes over the selected latitudinal band.  A second Fourier transform in time yields the frequency spectrum; the squared modulus 
$\lvert\widehat{v^{\prime}}\rvert^{2}$ is then filtered with a 31$-$point running median (along the frequency axis) to suppress broadband background power.  

For each of the three regions, the spectrum is Doppler-corrected to the local co-rotating frame, 
\begin{equation}
\omega^{\prime} \;=\; \omega - m\,\Delta\Omega,
\label{eq:doppler_shift}
\end{equation}
where $\omega^{\prime}$ is the intrinsic frequency, i.e. the frequency in the local co-rotating frame, and $\Delta\Omega(r,\theta)=\overline{\Omega}(r,\theta)-\Omega_{\rm frame}$  measures the difference between the local mean rotation and the angular velocity of the rotating frame \citep{2024ApJ...966...29B}, interpolated onto a uniform grid spanning $-500 \le \omega \le 200\ \mathrm{nHz}$, so that the Doppler-corrected frequencies, which are unevenly spaced for different $m$, are re-sampled on a common axis. The power spectra are visualized as $\log_{10}\!\bigl(P^{\,\Gamma}\bigr)$ with $\Gamma = 1.5$ (a mild gamma correction that enhances the narrow peaks of coherent modes).  

A common color scale is enforced for all panels, and the theoretical dispersion relation for equatorial and high-latitude modes is overplotted (see yellow lines) for the different radii and latitudes. For the high-latitude oscillations, analyzed in the local co-rotating frame, we adopt the approximation: $\omega^{\prime} \approx m\Delta\Omega_p$, where $\Delta\Omega_p$ denotes the difference in rotation between the polar regions ($55^\circ < \theta < 75^\circ$) and the local rotation at the latitude of the mode.

The diagonal “cut-off’’ visible in the several panels is an artifact of the analysis, produced by the combination of a strong Doppler correction and the finite frequency resolution of the time series (Nyquist frequency). At deeper layers and polar latitudes, the angular-velocity difference relative to the reference rotation rate is the largest, hence the Doppler shift displaces power increasingly far in frequency as \(m\) increases. No relevant physical information is lost in the portion of the spectrum that has been cut-off.

The spectra reveal that high-latitude modes are prevalent across all latitudes and depths, except in simulation DR10, where these modes do not appear in the stable layer. Within the frequency band explored in this analysis, it is evident that for shear greater than 20\%, the number of excited modes decreases with increasing shear, in agreement with the dispersion relation. In simulation DR20, modes with $m \lesssim 10$ are excited, whereas in DR40, only modes with $m \lesssim 5$ gain sufficient power.

At high and intermediate latitudes, the distribution of modes is more scattered. While some branches of excited modes follow the dispersion relation, others do not. The nearly horizontal branches around $\omega = 0$ correspond to modes that propagate at the same angular velocity as the local differential rotation. These are commonly referred to in the literature as critical latitude modes. Such modes are not observed at equatorial latitudes.

Equatorial Rossby modes do not appear to be significantly excited, or at least not with sufficient amplitude. Only a faint trace of them is loosely visible following the Rossby mode dispersion relation. We performed two additional numerical experiments to investigate the excitation of these modes by superimposing sectoral (DR20S) and tesseral (DR20T) perturbations on a snapshot of simulation DR20, taken in a statistically steady state, as initial condition. In DR20S, the perturbation corresponds to a spherical harmonic with $m = l = 10$, while in DR20T we used $m = 2$ and $l = 9$. The amplitude of the imposed perturbation is approximately $20$ m s$^{-1}$.

\begin{figure*}
\begin{center}
	\includegraphics[width=0.48\columnwidth]{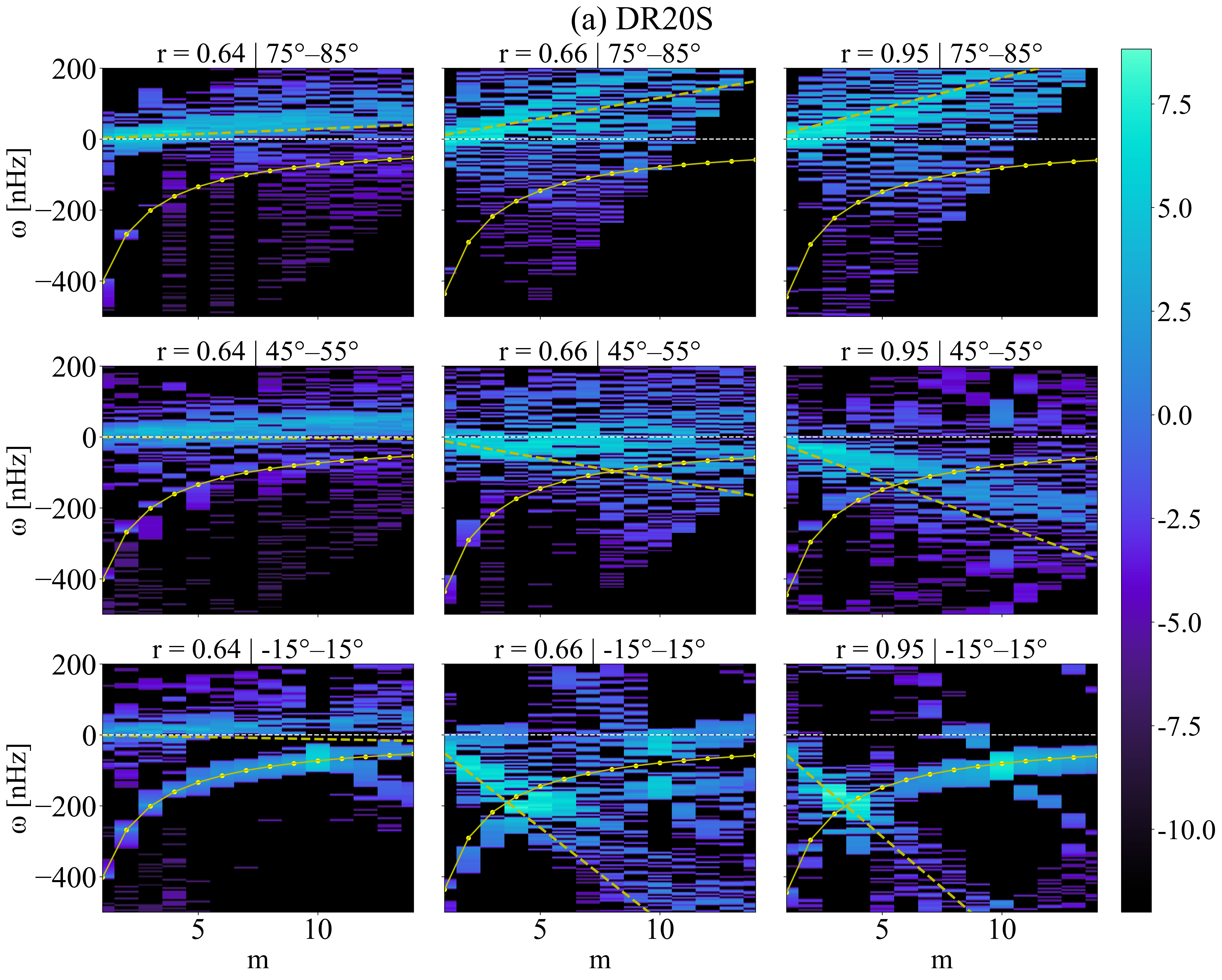} 
	\includegraphics[width=0.48\columnwidth]{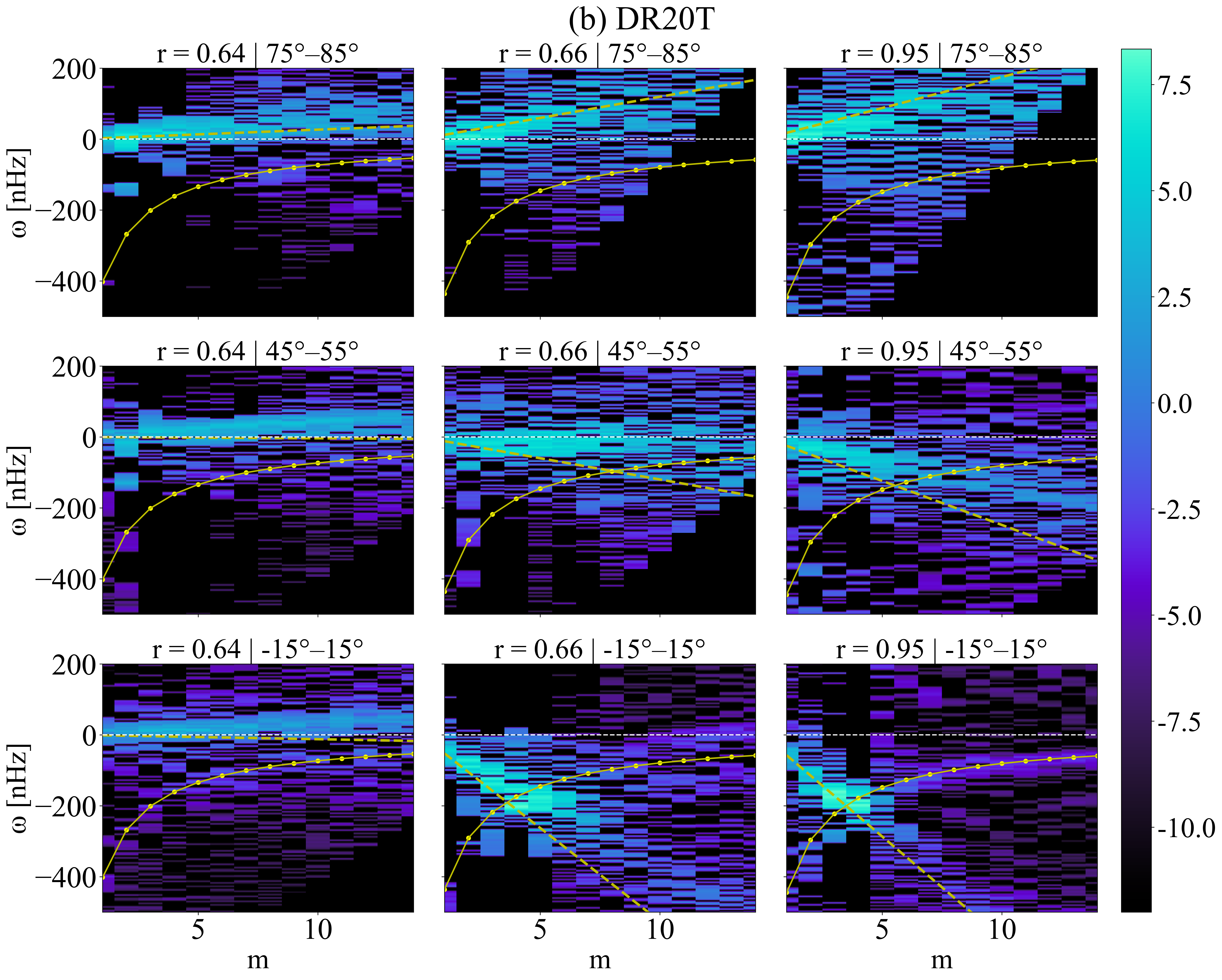}
    \caption{Similar to Fig.~\ref{fig:sp_shear} but for simulations DR20S (left) and DR20T (right).}
    \label{fig:sp_pert}
\end{center}
\end{figure*}

The resulting spectra of these simulations are shown in Fig.~\ref{fig:sp_pert} panels (a) and (b) for simulations DR20S and DR20T, respectively. The panels at the bottom rows, corresponding to equatorial latitudes, clearly display a Rossby mode branch roughly following the theoretical dispersion relation and with similar amplitude than the high latitude modes.  This confirms the findings discussed in \S\ref{s:solid} and previously reported by \citet{dikpati2022simulating}: even when a single, sectoral or tesseral, mode is initially excited, both, direct and inverse energy cascades transfer energy across the branch of Rossby modes. 

Notably, in Fig.~\ref{fig:sp_pert}(a), the $m=10$ mode stands out prominently, as it is the initial perturbation from which the others are excited. This behavior may offer a useful diagnostic tool for helioseismology, enabling the inference of the azimuthal wavenumber of the perturbations responsible for exciting Rossby modes within the solar convection zone.

\subsubsection{Angular momentum transport}

The compelling question regarding inertial modes is what role they play in the dynamics of the solar interior. What is their contribution to the angular velocity? How do they influence, or are they influenced by, the magnetic field? The results of this study provide insights into the first of these questions. To address it, we compute the angular momentum flux integrated over radial shells, $I_r$, and over conical surfaces with circumferences located at different latitudes, $I_{\theta}$. The expressions for $I_r$ and $I_{\theta}$ are given, for instance, in \citet[equations (14)–(16) of][]{guerrero2022implicit}.

\begin{figure*}
\begin{center}
    \includegraphics[width=0.48\columnwidth]{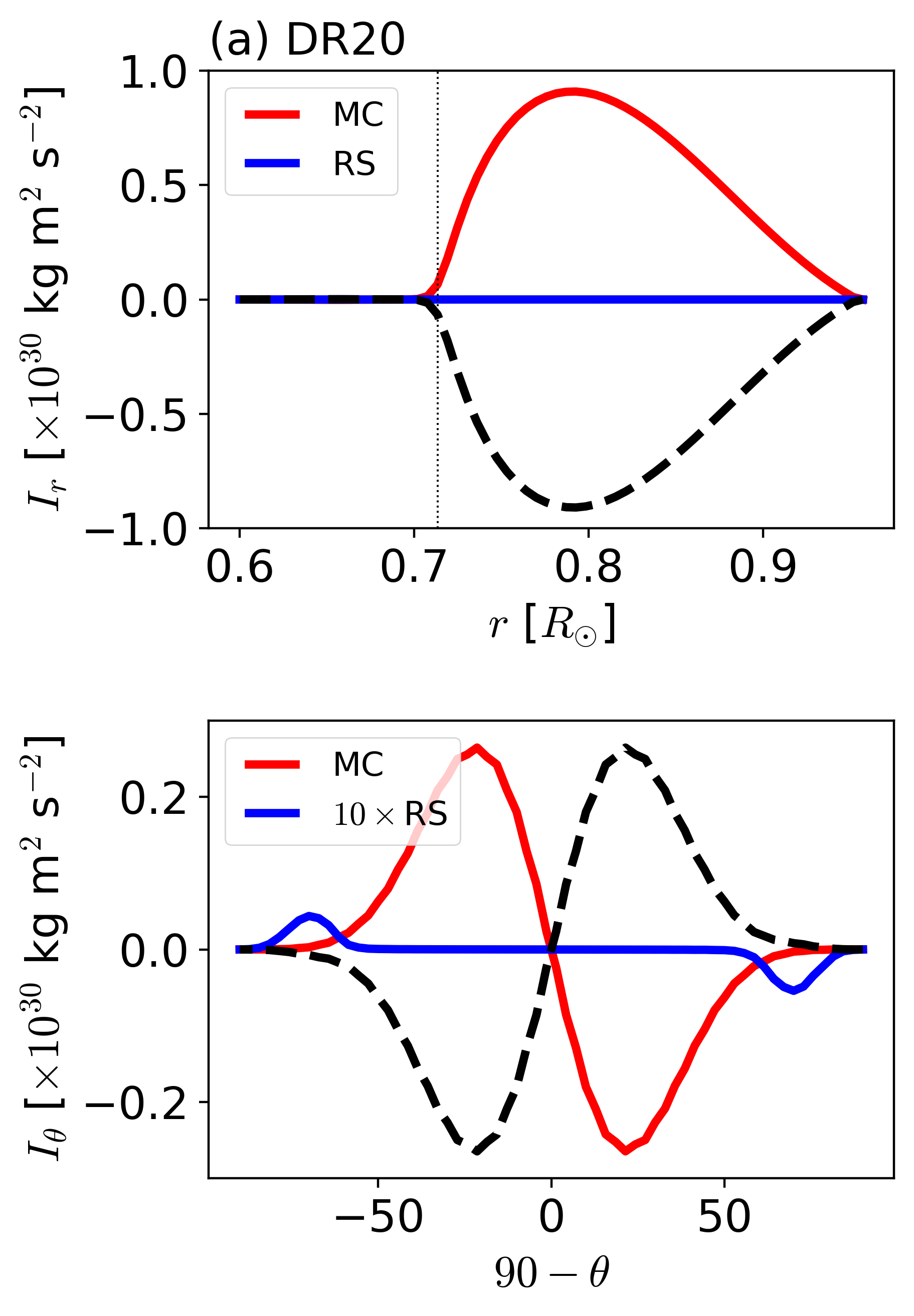}
    \includegraphics[width=0.48\columnwidth]{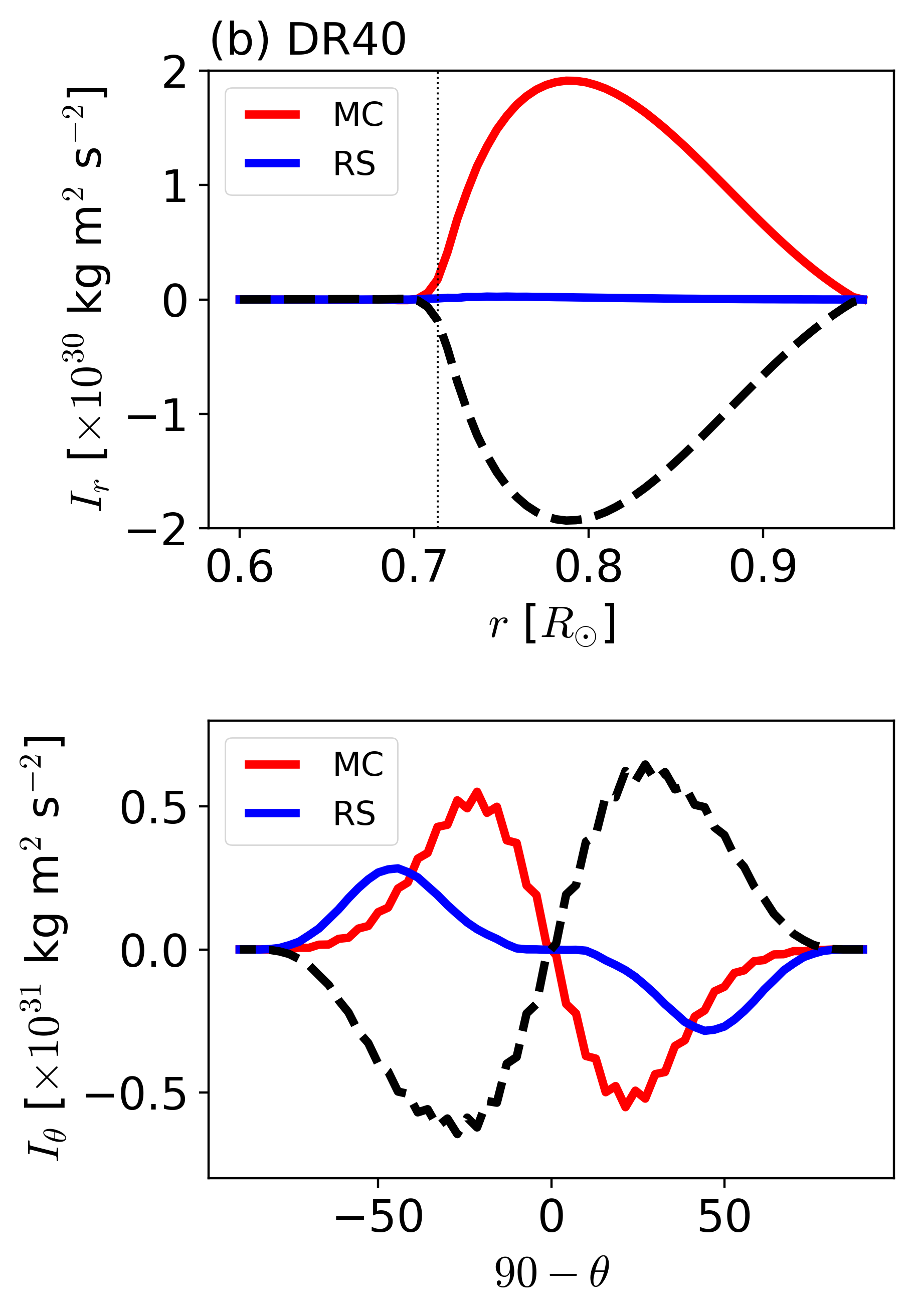}
    \caption{Integrated transport of angular momentum in radius (upper panels) and latitude (bottom panels)  by meridional circulation (red lines) and Reynolds stresses (blue lines) for simulations DR20 and DR40 at left and right, respectively. The dashed black line correspond to the residual of the MC and RS, which we interpret her as the contribution from the body force to the angular momentum flux. In the bottom-left panel the RS integrated flux was multiplied by a factor 10 to improve visibility.}
    \label{fig:ang_mom}
\end{center}
\end{figure*}

Figures~\ref{fig:ang_mom}(a) and (b) display $I_r$ (upper panels) and $I_{\theta}$ (lower panels) for simulations DR20 and DR40, respectively. The red and blue lines represent the flux contributions from meridional circulation and Reynolds stresses, respectively. The black dashed line corresponds to the residual between the two, which we interpret as the contribution from the body force term in Eq.\ref{eq:mom}. There should also be an angular momentum flux due to numerical dissipation; however, its effect should be negligible when compared to the amplitude of the forcing term. 

In the radial direction, we observe that the body force transports angular momentum downward. It is balanced by the net upward transport due to meridional circulation (MC). Reynolds stresses (RS) due to inertial modes play insignificant role in the radial transport of angular momentum. In latitude, the body force drives angular momentum toward the equator (positive values in the northern hemisphere), which is counterbalanced by meridional motions transporting angular momentum poleward—this advective flux is evident from the latitudinal velocity profiles in Fig.~\ref{fig:profs_dr}.

A key point in Fig.~\ref{fig:ang_mom} is the role of RS. For simulation DR20, their contribution is an order of magnitude smaller than that of MC (in the figure, the RS flux has been multiplied by a factor of $10$ for better visualization), indicating it is negligible. In contrast, for simulation DR40, the RS flux is comparable in magnitude to its MC counterpart and clearly contributes to balancing the transport induced by the body force. The negative values of the RS flux arise from the action of non-axisymmetric modes, which transport angular momentum poleward—thus accelerating the polar regions in opposition to the body force.

\begin{figure}
\begin{center}
	\includegraphics[width=0.8\columnwidth]{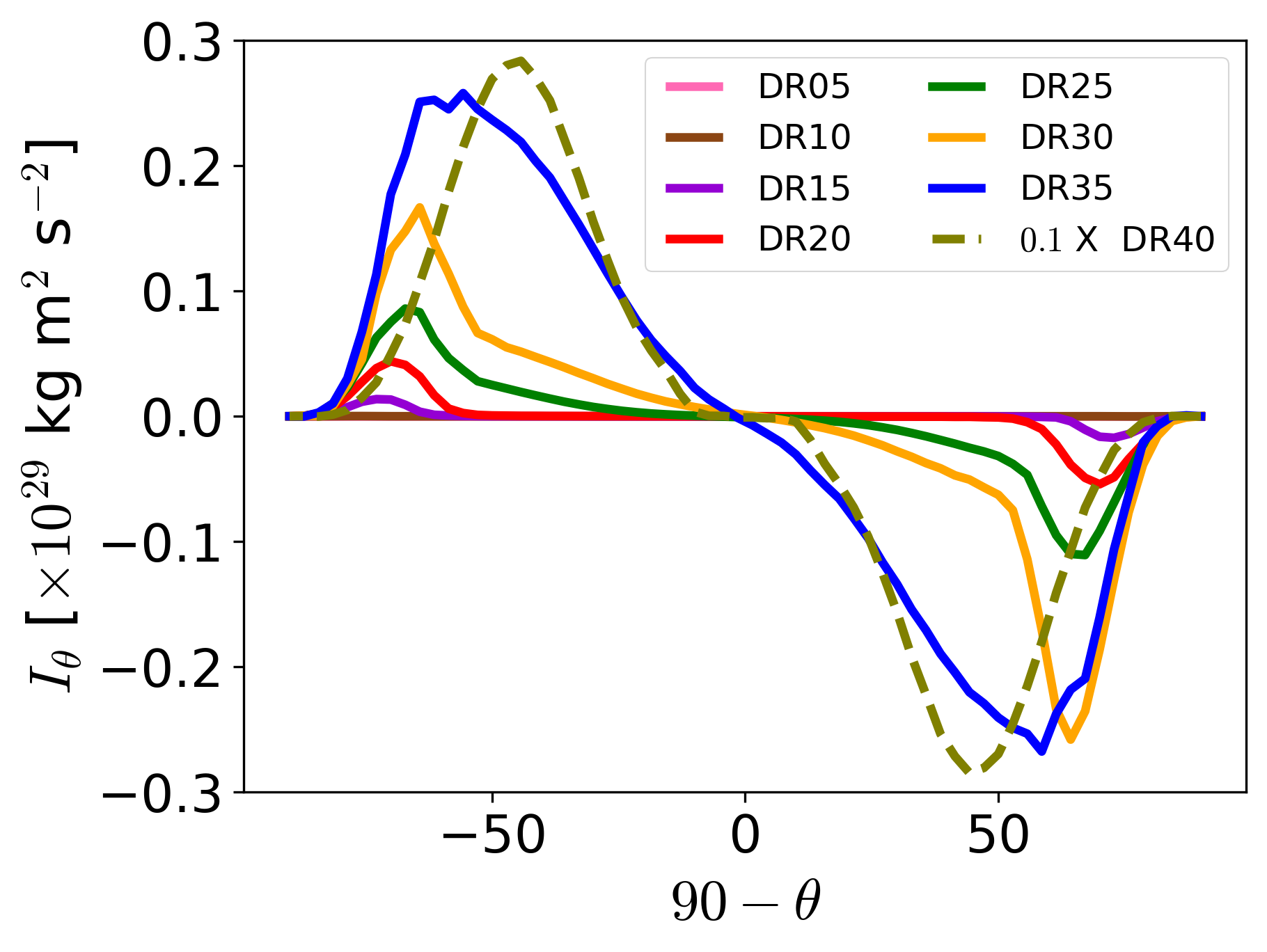}
    \caption{Integrated transport of angular momentum in latitude by Reynolds stresses.  The different colors correspond to different amount of forced shear according to the legend. The profile for 40\% shear is divided by a factor $10$ to improve presentation.}
    \label{fig:rs}
\end{center}
\end{figure}

Figure~\ref{fig:rs} further illustrates that the RS angular momentum flux increases with stronger shear. For weak shear, the flux peaks at polar latitudes, but as the shear increases, the peak shifts toward mid-latitudes. Notably, for a 40\% shear, the RS flux increases by roughly a factor of 10. Accompanying this angular momentum transport, there must also be a heat flux directed toward the equator, acting to homogenize the latitudinal temperature gradient imposed by thermal wind balance—particularly near the tachocline.

\section{Conclusions}
\label{s:conclusions}

We have presented a global anelastic hydrodynamic model that captures the excitation and nonlinear evolution of inertial waves in a solar-like, differentially rotating spherical shell. When excited with sectoral or tesseral spherical harmonics, the model self-consistently produces oscillatory modes that obey the Rossby-wave dispersion relation. Remarkably, non-linear energy cascades transfer energy towards both smaller and larger scales retaining frequencies always in agreement with the same dispersion relation. 

When differential rotation is imposed, with relative shear between 5\% and 40\%,  the system relaxes toward a state constrained by thermal wind balance, i.e., the imposed shear yields to the development of a latitudinal gradient of potential temperature as well as a meridional circulation velocity. It is worth noticing that a slightly sub-adiabatic thermal stratifications is required for obtaining tilted contours of rotation in the upper layer of our model.   Strong radial shear at tachocline depths triggers baroclinic instability, producing high-latitude inertial modes and polar vortices. The growth rate of these modes increases with shear and saturates at large amplitudes as the instability modifies the background shear. For a solar-like case, with 20\% shear, the dominant high-latitude modes exhibit frequencies comparable to those inferred from solar observations.

These baroclinically driven modes transport angular momentum poleward and heat equatorward, providing a plausible tachocline origin for the observed solar polar vortices.   This confirms, in global simulations, the results obtained by \cite{bekki2024sun} in mean-field models.   However, the contribution of the high latitude modes to the global differential rotation remains weak in our most solar-like cases, indicating that the Sun’s rotation profile may not be strongly modulated by these modes. Equatorial Rossby modes do not arise spontaneously but are readily excited by external perturbations, suggesting that their observed presence may reflect additional forcing from convective dynamics or magnetic fields. The absence of high-frequency modes further points to missing physics.

Taken together, our results support a tachocline-driven origin for high-latitude inertial modes while highlighting the need for magnetic fields and more realistic forcing to fully reproduce the observed solar inertial-wave spectrum and its dynamical impact.

\begin{acknowledgments}
We thank the anonymous referee for the constructive
comments and suggestions that have improved the quality of the paper.  
Computational resources supporting this work were provided by the NASA High-End Computing (HEC) Program through the NASA Advanced Supercomputing (NAS) Division at Ames Research Center. GG acknowledges financial support from CNPq. CFS and MSG acknowledge financial support from CAPES. This work is also supported by the NSF National Center for Atmospheric Research, which is a major facility sponsored by the National Science Foundation
(NSF) under cooperative agreement 1852977. MD acknowledges partial support from various NASA grants, such as NASA-HSR award 80NSSC21K1676 (awarded to NSF-NCAR), NASA-DRIVE Center award 80NSSC22M0162 to Stanford and NASA-HSR award 80NSSC21K1678 to JHU/APL. ST gratefully acknowledges funding from the European Research Council (ERC) under the European Union’s Horizon 2020 research and innovation program (Synergy grant agreement No. 855677 GRACEFUL).
\end{acknowledgments}

\newpage

\appendix

\section{Appendix A: Thermal wind balance equation}
\label{sec:apb}

The thermal wind equation is derived through the longitudinal component of the vorticity, which can be obtained from the curl of the momentum equation. Assuming a steady-state and an azimuthal average, the longitudinal component of the vorticity is given by:

\begin{equation}
\label{eq:bvt}  
\begin{split}
    \underbrace{2 \Omega_0 \frac{\partial \left \langle u_\phi \right \rangle }{\partial z}}_\textrm{Inertia} = &  \underbrace{- \left \langle \boldsymbol{\omega} \cdot \triangledown u_\phi  + \frac{\omega_\phi u_r}{r} + \frac{\omega_\phi u_\theta \cot(\theta)}{r} \right \rangle}_\textrm{Stretching}   \\
    & +   \underbrace{\left \langle   \boldsymbol{u} \cdot \triangledown \omega_\phi + \frac{u_\phi \omega_r}{r} + \frac{u_\phi \omega_\theta \cot(\theta)}{r} \right \rangle}_\textrm{Advection}   \\
    & +  \underbrace{\left \langle\omega_\phi \left [ \frac{1}{r^2}\frac{\partial(r^2 u_r) }{\partial r} +  \frac{1}{r\sin(\theta)}\frac{\partial (\sin \theta \ u_\theta) }{\partial \theta} +  \frac{1}{r\sin(\theta)}\frac{\partial u_\phi}{\partial \phi}   \right ] \right \rangle}_\textrm{Compressibility}    \\
    & +  \underbrace{\left \langle  \frac{g(r)}{r}\frac{\partial }{\partial \theta }\left ( \frac{\Theta^{\prime}}{\Theta_0} \right )  \right \rangle  }_\textrm{Baroclinicity}  \; .  
\end{split}
\end{equation}

Equation \ref{eq:bvt} is called Thermal Wind Balance and each term has a contribution to the meridional balance of the system.

\bibliographystyle{aasjournalv7}
\bibliography{references} 



\end{document}